\documentclass[sn-mathphys-ay,iicol]{sn-jnl}

\usepackage{graphicx}
\usepackage{amsmath,amssymb,amsfonts}
\usepackage{textcomp}
\usepackage{placeins}
\usepackage{etoolbox}
\usepackage{stfloats}
\usepackage{needspace}
\usepackage{capt-of}
\graphicspath{{image/}{figures/}}

\extrafloats{200}
\renewcommand{\dbltopfraction}{0.95}

\renewcommand{\dblfloatpagefraction}{0.8}

\makeatletter
\expandafter\let\csname sn@tablestar\expandafter\endcsname\csname table*\endcsname
\expandafter\let\csname endsn@tablestar\expandafter\endcsname\csname endtable*\endcsname
\renewenvironment{table*}[1][]%
{\csname sn@tablestar\endcsname[#1]%
\begin{center}%
\begin{threeparttable}%
\tablebodyfont%
\renewcommand\footnotetext[2][]{{\removelastskip\vskip3pt%
\let\tablebodyfont\tablefootnotefont%
\hskip0pt\if!##1!\else{\smash{$^{##1}$}}\fi##2\par}}%
}{\end{threeparttable}\end{center}\csname endsn@tablestar\endcsname}
\makeatother

\makeatletter
\patchcmd{\@maketitle}{\vskip21pt}{\vskip10pt}{}{\PackageWarning{fix}{t1 failed}}
\patchcmd{\@maketitle}{\removelastskip\vskip24pt}{\removelastskip\vskip16pt}{}{\PackageWarning{fix}{t2 failed}}
\patchcmd{\@maketitle}{\removelastskip\vskip24pt}{\removelastskip\vskip16pt}{}{\PackageWarning{fix}{t3 failed}}
\patchcmd{\@maketitle}{\removelastskip\vskip36pt}{\removelastskip\vskip20pt}{}{\PackageWarning{fix}{t4 failed}}
\makeatother

\makeatletter
\newenvironment{blocktable}
  {\par\addvspace{\intextsep}\begin{center}\begin{threeparttable}\tablebodyfont
   \renewcommand\footnotetext[2][]{{\removelastskip\vskip3pt%
   \let\tablebodyfont\tablefootnotefont%
   \hskip0pt\if!##1!\else{\smash{$^{##1}$}}\fi##2\par}}}
  {\end{threeparttable}\end{center}\par\addvspace{\intextsep}}
\newenvironment{blockfigure}
  {\par\addvspace{\intextsep}\begin{center}}
  {\end{center}\par\addvspace{\intextsep}}
\makeatother

\begin{document}
\title[Financial Literacy and Subjective Time Deformation]{Financial
Literacy and the Functional Shape of Subjective Time Deformation in
Intertemporal Choice}

\author*[1]{\fnm{Fabrizio} \sur{Maturo}}\email{fabrizio.maturo@unimercatorum.it}

\author[2]{\fnm{Salvador} \sur{Cruz Rambaud}}\email{scruz@ual.es}

\author[3]{\fnm{Vincenzo} \sur{Li Calzi}}\email{vincenzo.licalzi@studenti.unimercatorum.it}

\author[1]{\fnm{Andrea} \sur{Mazzitelli}}\email{a.mazzitelli@unimercatorum.it}

\author[4]{\fnm{Annamaria} \sur{Porreca}}\email{annamaria.porreca@uniroma5.it}

\affil*[1]{\orgdiv{Department of Economics, Statistics and Business,
Faculty of Technological \& Innovation Sciences},
\orgname{Universitas Mercatorum},
\orgaddress{\city{Rome}, \country{Italy}}}

\affil[2]{\orgdiv{Departamento de Econom\'ia y Empresa},
\orgname{Universidad de Almer\'ia},
\orgaddress{\city{Almer\'ia}, \country{Spain}}}

\affil[3]{\orgdiv{Department of Engineering and Science},
\orgname{Universitas Mercatorum},
\orgaddress{\city{Rome}, \country{Italy}}}

\affil[4]{\orgdiv{Department for the Promotion of Human Science and
Quality of Life}, \orgname{San Raffaele University},
\orgaddress{\city{Rome}, \country{Italy}}}

\abstract{
Intertemporal decisions are commonly summarized by a single discount parameter, although individuals with similar overall discounting intensity may value short-, medium-, and long-term delays differently. This study investigates whether financial literacy is associated with these horizon-specific differences by treating reconstructed subjective time as a functional rather than scalar outcome. Data were collected through a multilingual online questionnaire administered to a convenience and snowball sample of 140 adults, who provided hypothetical equivalence judgments for a future reward of EUR 100 over 13 horizons from 7 days to 5 years, a three-item financial-literacy measure, and measures of temporal perception. Individual discount trajectories were reconstructed and analyzed with function-on-scalar regression, allowing the association with financial literacy to vary across the temporal domain. Because the normalized subjective-time trajectory is not identifiable for flat or reconstruction-degenerate discount profiles, the primary functional analysis included 59 participants, while complementary analyses used the full analytical sample. The primary model provided no evidence of a robust linear association between financial literacy and subjective-time deformation. An exploratory categorical specification indicated heterogeneity across financial-literacy levels and departure from a linear-score representation, although individual contrasts were imprecise and sensitive to specification. The relationship between financial literacy and intertemporal valuation may thus be more complex than a monotonic score effect, and functional data analytics can reveal horizon-specific structure that scalar summaries obscure. Given the convenience sample and the reduced identified sample, the results are exploratory and require validation in larger independent samples.
}


\maketitle

\section{Introduction}
\label{sec:intro}

Intertemporal choices describe decision situations in which the positive and negative consequences of the considered alternatives emerge at different points in time. The classical discounted utility model assumes that the value of future outcomes decreases according to an exponential discount function and describes individual preferences through a scalar discount parameter \citep{Samuelson_1937}. However, this model is unable to describe some behaviors observed in actual choices, such as preference reversals, present bias, and decreasing impatience, which are not represented by a model based on a constant discount rate \citep{Loewenstein_1989, Loewenstein_1992}. One possible explanation is that individuals assign value to future outcomes not only by considering their objective temporal location, but also on the basis of a subjective representation of temporal distance \citep{Cruz_Rambaud_2023, Zauberman_2009}.

From this perspective, the discounting process can be represented through an exponential structure in which chronological time is transformed differently for each individual. The resulting function \(\tau_i(t)\) describes how each individual reorganizes chronological time in their intertemporal choices. Treating this trajectory as a functional object, reconstructed from intertemporal equivalence judgments, makes it possible to study individual differences through the characteristics of its shape \citep{maturo2026subjectivetimedeformationintertemporal}. Functional data analysis entered this field through the representation of the discount function itself as a functional object, which has been used to cluster behavioural investor profiles \citep{Maturo_2026_fclust}, to characterize anomalies in intertemporal decisions \citep{Ventre_2024_hikikomori}, and to compare different dimensions of the discount function \citep{Martino_2025_dimensions}. The present study differs in two respects: the functional object is the reconstructed subjective-time trajectory rather than the discount function, and it enters a regression model as the response, so that the association with an individual covariate is estimated across the temporal domain rather than described. A descriptive functional analysis of this kind, however, does not establish whether a specific individual covariate is associated with the shape of the trajectory. The present study examines this relationship by considering
\begin{equation}
\label{eq:tau_centered}
\tau_i^c(t)=\tau_i(t)-\bar{\tau}(t)
\end{equation}
as a functional response and by studying its association with financial literacy across the entire temporal domain. Financial literacy plays an important role in this context, since many financial decisions require an evaluation between outcomes that occur over different time horizons. The literature has shown associations between financial knowledge, saving, wealth accumulation, retirement planning, and other decisions whose effects unfold over time \citep{LUSARDI_2011,Lusardi_2014}. At the same time, financial literacy cannot be considered a direct measure of rationality or an indicator of optimal behavior, since it includes a plurality of dimensions, including skills, attitudes, and behaviors, that cannot be described through measures of financial knowledge alone \citep{HUSTON_2010}.

The objective of the study is to evaluate the possible relationship between financial literacy and the functional form of implicit subjective time deformation. The aim is not to estimate a causal effect of financial literacy nor to assume that higher values correspond to greater individual rationality. Financial literacy is considered as a scalar covariate that may be associated with the configuration of subjective temporal distance over the observed horizon. The functional response is represented by the centered trajectory \(\tau_i^c(t)\), while the Temporal Sense Scale and age are considered as control covariates. Since the reconstruction jointly depends on \(k_i\) and \(\tau_i(t)\) through the product \(k_i\tau_i(t)\), a formulation including \(\log(k_i)\) is also considered in order to evaluate the stability of the coefficient associated with financial literacy after accounting for individual discounting intensity.

A valid discount-curve reconstruction was obtained for \(N=139\) participants. As detailed in Section~\ref{sec:methods}, \(\tau_i(t)\) is mathematically unidentified for participants whose raw discounting profile is exactly flat, and it is operationally unidentified for raw non-flat participants whose monotone reconstruction yields a numerically negligible discounting scale. The primary function-on-scalar regression is therefore estimated on the \(N=59\) participants with an identified reconstructed trajectory, and the full sample flow is reported in Table~\ref{tab:sample_description}. Exact-flat status is modeled separately on the full sample, and a complementary function-on-scalar regression uses \(g_i(t)=-\log\widehat f_i(t)\), which is defined for all \(139\) participants. Scalar models based on \(k_i\), \(\log(k_i)\), and the leading FPCA scores provide additional benchmarks \citep{Ramsay_2005}. Sensitivity analyses evaluate the consequences of alternative financial-literacy coding, the reconstruction-identification tolerance, near-flat thresholds, smoothing specifications, functional outliers, educational attainment, the inclusion of \(\log(k_i)\), and an alternative function-on-scalar estimator. The reconstructed trajectory is interpreted throughout as an implicit, normalized trajectory conditional on the individual reference scale \(k_i\), rather than as a direct psychophysical measure of subjective time.

Financial literacy was included in the questionnaire, together with other covariates, on the basis of the literature on intertemporal choice, which has already used this variable. Its association with the functional configuration of the reconstructed subjective time trajectory was formulated during the exploratory functional analysis rather than as a pre-specified hypothesis. The analysis is therefore a secondary exploratory analysis, and the corresponding results are preliminary evidence useful for formulating hypotheses to be evaluated in independent samples.

The remainder of the paper is organized as follows. Section~\ref{sec:background} develops the conceptual background linking financial literacy, intertemporal choice, and functional subjective time. Section~\ref{sec:data} describes the data, measures, and construction of the functional response. Section~\ref{sec:methods} presents the statistical methods, including the identification of flat discounting profiles and the resulting two-part analytical strategy. Section~\ref{sec:results} reports the flat-profile diagnostic evidence, descriptive functional profiles, regression results on the identified sample, the complementary full-sample analysis, and sensitivity and robustness analyses. Section~\ref{sec:discussion} discusses the interpretation of the findings. Sections~\ref{sec:limitations} and \ref{sec:conclusions} present the limitations and conclusions, respectively. Additional diagnostic tables and robustness checks are reported in the Appendix.

\section{Conceptual Background}
\label{sec:background}

In intertemporal choice, individuals are required to make decisions whose effects are not concentrated at a single point in time but are instead distributed over a more or less extended time horizon. This aspect is particularly relevant in finance, where many economic decisions involve the need to balance immediate gratification with future outcomes, as in the case of saving, investment, borrowing, and retirement planning \citep{Ventre_2024_b}. The classical discounted utility model describes intertemporal choice on the basis of the assumption that the weight assigned to future outcomes decreases exponentially with temporal distance, thereby implying a constant discount rate and temporally consistent preferences \citep{Samuelson_1937}. However, subsequent literature has shown that the exponential model does not always adequately describe observed behavior. A number of contributions have documented a set of behavioral anomalies, highlighting recurring phenomena such as present bias, preference reversals, and decreasing impatience \citep{Frederick_2002}. Studies have explained these anomalies by hypothesizing that individuals do not evaluate time as an objective and linear dimension, but rather represent it subjectively during the evaluation of alternatives \citep{Loewenstein_1992,Zauberman_2009}. However, deviations from the model cannot be interpreted as mere decision errors or as evidence of individual irrationality. One possible interpretation is that time is not evaluated exclusively in its objective dimension but also through the subjective representation constructed by each individual. Consequently, the way in which a future reward is evaluated depends on the magnitude of the reward itself, the temporal delay, and the perception of waiting relative to the present \citep{Cruz_Rambaud_2023}.

In this context, financial literacy can be considered particularly relevant because it concerns individuals' ability to understand financial information and to use such information in decisions that have medium and long-term effects. One strand of the literature has shown that financial literacy is associated with relevant financial behaviors, such as wealth accumulation, retirement planning, and savings management \citep{Lusardi_2007,Lusardi_2014}. A second strand has linked low levels of financial knowledge to problematic decisions in specific domains, such as borrowing and credit use \citep{LUSARDI_2015}. These results suggest the importance of including financial literacy in the analysis of intertemporal choice, without however allowing it to be interpreted as an equivalent of rationality. One reason for maintaining such caution lies in the fact that financial literacy does not coincide with financial knowledge alone, but also includes behavioral and attitudinal components, commonly identified as financial behavior and financial attitude. Knowledge of financial concepts may foster greater awareness in decision-making, but it is not sufficient to guarantee appropriate behavior or fully consistent management of resources. It is therefore possible for an individual to exhibit a high level of financial competence while at the same time making choices or expressing attitudes that are not fully consistent with such competence. For this reason, decision quality depends not only on knowledge considered in isolation, but also on the interaction between knowledge, behavior, and attitudes \citep{Ventre_2024_b,OECD2026Toolkit}. \citet{Bawalle_2024} adopted the distinction between financial knowledge, financial behavior, and financial attitude, relating it to hyperbolic discounting, and reported that the three components show a negative association with the tendency to prefer immediate rewards over future ones. That result is an empirical relationship between components of financial literacy and intertemporal preferences, and not evidence that more financially literate individuals are necessarily more rational in a general sense.

A second reason for caution concerns the direction of the relationship between financial literacy and time preferences. One strand of the literature interprets the acquisition of financial information as a form of human capital investment, since it requires a present cost in view of future benefits \citep{Becker_1997}. Indeed, time preferences may influence the likelihood that an individual undertakes financial education, as those who place strong weight on the present tend to be less likely to voluntarily participate in training programs, even when participation costs are limited \citep{Meier_2013}. Consequently, the relationship between financial literacy and intertemporal choice should not be conceived as necessarily unidirectional. However, results from experimental studies show that financial education can reduce forms of temporal inconsistency, increase consistency with the law of demand, and affect the way individuals allocate monetary rewards across different points in time \citep{L_hrmann_2018}. Another strand interprets delay discounting as a behavioral measure of decisional impulsivity, in which the subjective value of a reward decreases as delay increases, and steeper discounting curves indicate a stronger preference for immediate rather than future outcomes \citep{Mazur1987AnAP}. In this context, financial education has been associated with a reduction in delay discounting, suggesting a possible effect on the evaluation of future rewards \citep{DeHart_2016}. However, this effect must be distinguished from other factors such as framing and temporal representation, which can alter the valuation of delayed rewards \citep{Peters_2010,DeHart_2014}. In this sense, greater financial literacy may affect task interpretation rather than underlying impatience. Some studies based on monetary time-preference elicitation tasks show that individuals with higher financial literacy are more likely to take into account, in their evaluations, borrowing and saving opportunities available outside the experimental setting. For this reason, observed patience may be interpreted as the result not only of individual time preferences, but also of mechanisms related to broad bracketing or intertemporal arbitrage \citep{Oberrauch_2022}. \citet{Zauberman_2009} suggested that the subjective perception of future duration does not necessarily follow a linear pattern with respect to objective time, but may instead take a concave form, which provides a possible explanation for empirically observed hyperbolic discounting patterns. \citet{Kim_2009} distinguished two dimensions of anticipatory time perception, the first concerning diminishing sensitivity associated with longer time horizons and the second the overall level of time contraction, and reported that both components play a relevant role in temporal discounting processes. In the present work, financial literacy is considered a central covariate not because it constitutes a direct measure of rationality, but because it can be theoretically linked to the evaluation of the future, planning abilities, the understanding of intertemporal trade-offs, and the management of financial uncertainty.

The central methodological distinction concerns the scalar discounting scale \(k_i\) and the reconstructed functional trajectory \(\tau_i(t)\). Within the adopted operational reconstruction, \(k_i\) provides an individual reference scale for discounting, while \(\tau_i(t)\) describes the functional configuration obtained after normalizing the smoothed discount trajectory by this scale. The individual discount curve is represented as
\begin{equation}
\label{eq:discount}
f_i(t)=\exp\{-k_i\tau_i(t)\},
\end{equation}
and, once \(k_i\) has been estimated according to the normalization procedure described in Section~\ref{sec:methods}, the implicit subjective time trajectory is reconstructed as
\begin{equation}
\label{eq:tau}
\tau_i(t)=-\frac{\log f_i(t)}{k_i}.
\end{equation}
Accordingly, \(k_i\) and \(\tau_i(t)\) are treated as distinct operational components of the reconstruction, but this decomposition should not be interpreted as establishing a unique psychophysical separation between discount intensity and subjective time. A limiting case arises when a participant's observed discounting curve shows no reduction in the immediate equivalent of the future reward at any elicited horizon. Then \(f_i(t)=1\) at every observed \(t\), so that \(-\log f_i(t)=0\) at every observed \(t\) and the discount intensity is \(k_i=0\). The ratio defining \(\tau_i(t)\) is then \(0/0\), an indeterminate form under which any trajectory is compatible with the observed data. For such participants \(\tau_i(t)\) is therefore mathematically unidentified rather than weakly identified. This distinction motivates the two-part strategy adopted in Section~\ref{sec:methods}. The function \(\tau_i(t)\) is therefore interpreted, for the participants for whom it is identified, as an implicit reconstructed and normalized subjective time trajectory conditional on the individual reference scale \(k_i\). The centered version \(\tau_i^c(t)\) in Equation~\eqref{eq:tau_centered} removes the sample mean trajectory pointwise and allows individual functional deviations from the mean trajectory to be examined across the temporal domain, without implying a complete separation between functional shape and individual amplitude or scale. This approach is useful because two individuals may exhibit similar values of \(k_i\), and therefore a seemingly comparable overall discounting intensity, but differ in the shape of the trajectory \(\tau_i^c(t)\). Such differences may concern specific regions of the time horizon, such as the short, medium, or long term, and may reflect different forms of compression, expansion, or subjective reorganization of temporal distance.

The contribution of this study lies in examining the possible association between financial literacy and the functional structure of implicit subjective time, represented by the trajectory \(\tau_i^c(t)\). The analysis concerns the possible association between levels of financial literacy and different configurations of subjective temporal distance, with the aim of enriching the understanding of the relationships between financial skills, valuation of the future, and intertemporal choices.

\section{Data and Measures}
\label{sec:data}

The study is based on a multilingual online questionnaire designed to elicit intertemporal equivalence judgments, measures of temporal perception, and individual characteristics potentially associated with the evaluation of the future. The analysis focuses on the centered functional trajectory of implicit subjective time, the financial-literacy score, the Temporal Sense Scale, age, and the individual discount-intensity parameter used in sensitivity analyses. The questionnaire includes intertemporal equivalence judgments over thirteen time horizons ranging from 7 days to 5 years, through which each participant reports the immediate amount considered equivalent to a fixed future reward.

The intertemporal equivalence task was hypothetical. A fixed future amount of \(100\) euros was used as the reference value, and no financial incentives or other compensation were linked to the elicitation task. Data collection took place between February and May 2026. Participants were recruited through non-probability convenience and snowball dissemination of the survey link through personal contacts, contacts of initial recipients, and social media platforms. The sample was not designed to be representative of a defined population, and all empirical results are interpreted as sample-specific exploratory evidence. The questionnaire was available in Italian, English, and Spanish. No formal back-translation or independent linguistic validation procedure was conducted. Participation was voluntary, and participants were informed about the scientific purpose of the study and the intended use of the collected data for scientific and academic research.

The raw dataset contains \(140\) participants. One participant provided only \(4\) valid equivalence observations across the \(13\) elicited horizons and did not meet the minimum requirement of \(6\) valid elicited observations for individual discount-curve smoothing. A valid discount-curve reconstruction and complete values of financial literacy, the Temporal Sense Scale, and age were available for \(139\) participants. Of these, \(61\) had an exact-flat raw discounting profile and \(78\) were raw non-flat. Among the raw non-flat participants, \(19\) had a reconstruction-degenerate profile after monotone smoothing, defined by \(k_i<10^{-8}\); the primary \(\tau_i^c(t)\) analysis therefore includes \(59\) participants with an identified reconstructed trajectory. The main sample flow is reported in Table~\ref{tab:sample_description}.

\begin{table*}[!htb]
\centering
\caption{Sample description and analytical flow (\(N=139\) full analytical sample). Exact-flat status is defined from the raw equivalence judgments. The reconstruction-degenerate category consists of raw non-flat participants for whom the monotone reconstructed discount curve yields \(k_i<10^{-8}\). The distribution of \(k_i\) is therefore summarized only for the \(N=59\) primary identified sample.}
\label{tab:sample_description}
\small
\begin{tabular}{@{}llr@{}}
\toprule
\textbf{Characteristic} & \textbf{Category} & \textbf{Value} \\
\midrule
\multicolumn{3}{@{}l}{\textit{Panel A -- Sample flow}} \\
Initial respondents & & 140 \\
Valid discount-curve fit & & 139 \\
Full analytical sample & & 139 \\
\quad Raw exact-flat & & 61 \\
\quad Raw non-flat & & 78 \\
\quad Reconstruction-degenerate raw non-flat (\(k_i<10^{-8}\)) & & 19 \\
\quad Identified \(\tau_i(t)\), primary functional sample & & 59 \\
\midrule
\multicolumn{3}{@{}l}{\textit{Panel B -- Demographics}} \\
Gender & Female & 77 (55.4\%) \\
       & Male & 62 (44.6\%) \\
Age (years) & Mean (SD) & 39.6 (12.1) \\
            & Range & 18--73 \\
Education & Middle school & 2 (1.4\%) \\
          & High school & 42 (30.2\%) \\
          & Bachelor's degree & 23 (16.5\%) \\
          & Master's degree & 57 (41.0\%) \\
          & Doctorate & 12 (8.6\%) \\
          & Other & 3 (2.2\%) \\
Income & \({<}\,\){\texteuro}10,000 & 27 (19.4\%) \\
       & {\texteuro}10,000--{\texteuro}20,000 & 27 (19.4\%) \\
       & {\texteuro}20,001--{\texteuro}40,000 & 40 (28.8\%) \\
       & {\texteuro}40,001--{\texteuro}60,000 & 12 (8.6\%) \\
       & {\texteuro}60,001--{\texteuro}100,000 & 5 (3.6\%) \\
       & Prefer not to say & 28 (20.1\%) \\
Language & Italian & 130 (93.5\%) \\
         & English & 5 (3.6\%) \\
         & Spanish & 4 (2.9\%) \\
Nationality & Italian & 134 (96.4\%) \\
            & Other & 5 (3.6\%) \\
\midrule
\multicolumn{3}{@{}l}{\textit{Panel C -- Main variables}} \\
Financial literacy & \(FL = 0\) & 11 (7.9\%) \\
                   & \(FL = 1\) & 39 (28.1\%) \\
                   & \(FL = 2\) & 55 (39.6\%) \\
                   & \(FL = 3\) & 34 (24.5\%) \\
TSS & Mean (SD) & 4.86 (1.15) \\
    & Range & 1.00--7.00 \\
\(k_i\), identified primary sample & Median & \(1.33\times10^{-4}\) \\
      & Range & \([9.43\times10^{-7},\,1.02\times10^{-2}]\) \\
\botrule
\end{tabular}
\end{table*}

For each individual \(i\) and each time horizon \(t\), the equivalence judgment is transformed into a value of the individual discounting function. Denoting by \(A_{0,i}(t)\) the immediate amount considered equivalent to the fixed future reward \(A_f=100\), the observed discounting curve is defined as
\begin{equation}
f_i(t)=\frac{A_{0,i}(t)}{A_f}.
\end{equation}
Lower values of \(f_i(t)\) indicate a stronger devaluation of the future reward, whereas values closer to one indicate a smaller subjective reduction in future value.

Since the values of \(f_i(t)\) are observed only on a finite set of time horizons, they are treated as discrete points belonging to an underlying trajectory. Functional Data Analysis allows these observations to be treated as discrete realizations of a function defined on a continuous domain, shifting the statistical unit of analysis from the single observed value to the entire curve \citep{Ramsay_2005}. This approach allows both discounting curves and subjective time functions to be represented as individual trajectories, preserving the information contained in their shape along the observed time horizon.

The functional reconstruction phase requires a preliminary regularization step, since equivalence judgments may contain measurement noise, missing observations, or small local inconsistencies with respect to the theoretical shape of the discounting curve. For this reason, values outside the domain of the discounting function, namely \(f_i(t)\leq 0\) or \(f_i(t)>1\), are treated as missing, while the initial value \(f_i(0)=1\) is imposed as a theoretical constraint. Subsequently, for each individual, the curve \(f_i(t)\) is estimated through a monotonically decreasing smoothing procedure, in order to satisfy the principle that the subjective value of a future reward should not increase with waiting time. The reconstruction of implicit subjective time then applies Equations~\eqref{eq:discount} and \eqref{eq:tau}: once the individual reference scale \(k_i\) has been fixed, the preliminary subjective time trajectory is obtained from the smoothed discount function and is subsequently regularized through a monotonically increasing smoothing procedure. This second smoothing step enforces the monotonicity of the reconstructed trajectory and reduces local irregularities inherited from the empirical equivalence judgments. It should therefore be interpreted as a regularization step applied to the reconstructed trajectory rather than as a source of additional independent information. After this regularization, the common origin \(\tau_i(0)=0\) is imposed, yielding an implicit reconstructed and normalized subjective time trajectory conditional on the individual reference scale \(k_i\).

The main functional response is not the raw trajectory \(\tau_i(t)\), but its centered form \(\tau_i^c(t)\) defined in Equation~\eqref{eq:tau_centered}, where \(\bar{\tau}(t)\) denotes the sample mean function. This specification is motivated by the objective of examining individual functional deviations from the sample mean trajectory. Centering removes the sample mean function pointwise and yields individual deviations from the mean trajectory, but it does not by itself fully separate functional shape from individual amplitude or scale.

The main predictor is financial literacy, measured through a discrete score ranging from \(0\) to \(3\). The score is constructed from three financial-knowledge items concerning interest accumulation, inflation and purchasing power, and risk diversification. Each item contributes one point when answered correctly and zero otherwise, so that incorrect responses and ``I do not know'' responses receive the same score of zero. The resulting score therefore represents the number of correctly answered financial-knowledge items. This approach is consistent with the literature on brief financial literacy measures, designed to capture basic financial knowledge using a limited number of simple, relevant, short, and comparable questions \citep{LUSARDI_2011}. \citet{HUSTON_2010} highlights that financial literacy measurement faces standardization limitations and requires careful consideration of construct definition, item content, measurement procedure, and scoring. All four levels of the financial literacy score are represented in the primary identified sample (\(FL=0\): \(n=5\); \(FL=1\): \(n=11\); \(FL=2\): \(n=23\); \(FL=3\): \(n=20\)). In the main specification, financial literacy is treated as a discrete ordered numerical score ranging from \(0\) to \(3\), so that the corresponding coefficient function represents the average functional association associated with a one-point increase in the score. The adequacy of this parsimonious specification is evaluated through a formal comparison with a four-level categorical representation of financial literacy. In robustness analyses, the same variable is also recoded into three groups, low, mid, and high, where the low group includes levels \(0\) and \(1\), the mid group corresponds to level \(2\), and the high group corresponds to level \(3\).

The score used in the present study should be interpreted as a brief knowledge-based measure. It does not represent a comprehensive measure of financial literacy in a broad sense, since international measurement instruments also include components related to skills, attitudes, behaviors, and financial awareness. In its broader definition, financial literacy consists of a combination of awareness, knowledge, skills, attitudes, and behaviors necessary for making informed financial decisions and improving individual well-being \citep{OECD2026Toolkit}. \citet{KNOLL_2012} show that financial knowledge can also be measured using broader scales and Item Response Theory models, in which items are evaluated according to their difficulty and informativeness.

In addition to financial literacy, the model includes two control covariates, the Temporal Sense Scale and age. The Temporal Sense Scale is based on five items assessing the perceived speed of time with reference to everyday life, the week, the month, the year, and life as a whole, using a seven-point response scale ranging from very slow to very fast \citep{Zhang_2022}. In the present study, the five core items are summarized through their individual mean, yielding a score ranging from \(1\) to \(7\), with higher values indicating a faster perceived passage of time. The internal consistency of the five-item score in the analytical sample is satisfactory, with Cronbach's \(\alpha=0.879\). The Temporal Sense Scale is included because it represents a subjective dimension of temporal perception distinct from financial literacy and potentially associated with the structure of the trajectory \(\tau_i^c(t)\). Age is included as a demographic control covariate, since the representation of the future and the perception of temporal distance may vary across the life course. The parameter \(k_i\), estimated in the reconstruction phase of \(\tau_i(t)\), is not included in the main specification, as it represents individual discount intensity and is not the direct focus of the analysis. However, it is considered in a sensitivity specification in the form \(\log(k_i)\). The logarithmic transformation is used because the distribution of \(k_i\) is strongly skewed and the logarithmic scale provides a more stable representation in the sensitivity specification.

In summary, the analysis uses the centered trajectory \(\tau_i^c(t)\) as the functional response, the discrete financial literacy score ranging from \(0\) to \(3\) as the main predictor, the Temporal Sense Scale and age as control covariates, and \(\log(k_i)\) as a sensitivity variable. The primary specification addresses whether financial literacy is associated with functional deviations in the reconstructed subjective time trajectory along the observed temporal horizon; the sensitivity specification including \(\log(k_i)\) addresses whether such an association persists once the scalar intensity of discounting is accounted for. The parameter \(k_i\) and the trajectory \(\tau_i^c(t)\) are therefore treated as distinct operational components of the analysis, while recognizing that the reconstructed trajectory remains conditional on the individual reference scale \(k_i\).

\section{Statistical Methods}
\label{sec:methods}

\subsection{Functional Reconstruction and Identification}
\label{sec:methods_identification}

For each participant \(i\), the intertemporal equivalence judgments are converted into discount factors \(f_i(t)\), and the reconstructed subjective-time representation follows Equation~\eqref{eq:discount}. The observed discount factors are augmented with the theoretical anchor \(f_i(0)=1\). Participants are required to have at least six valid observations among the thirteen elicited horizons; the anchor at \(t=0\) is not counted toward this minimum.

The individual discount curve is estimated with a monotone decreasing shape-constrained additive model using the \texttt{mpd} basis in \texttt{scam} with basis dimension \(k=10\). The fitted curve is then divided by its fitted value at \(t=0\), thereby imposing \(\widehat f_i(0)=1\) numerically as well as through the anchor observation. Fitted values are constrained to the interval \([10^{-6},1]\). This renormalization matters for profiles with very weak fitted discounting, because a small fitted offset at the origin can otherwise dominate the subsequent ratio used to reconstruct \(\tau_i(t)\).

For raw non-flat participants, the individual discounting scale \(k_i\) is estimated as the ordinary least-squares slope through the origin in the regression of \(-\log\widehat f_i(t)\) on chronological time. A preliminary normalized trajectory is then obtained as
\begin{equation}
\tau_{raw,i}(t)=-\frac{\log\widehat f_i(t)}{k_i}.
\end{equation}
The preliminary trajectory is regularized with a monotone increasing \texttt{scam} smooth using the \texttt{mpi} basis with basis dimension \(k=8\), evaluated on a common grid of \(200\) equally spaced points from \(0\) to \(1825\) days. The fitted value at \(t=0\) is subtracted and residual negative numerical values are set to zero. The resulting \(\tau_i(t)\) is an implicit reconstructed and normalized trajectory conditional on the estimated individual reference scale \(k_i\), rather than a direct psychophysical measure of perceived duration.

Identification is assessed at two stages. First, an exact-flat participant is one for whom the raw equivalence amount equals the fixed future reward at every elicited horizon, implying \(f_i(t)=1\), \(k_i=0\), and an unidentified ratio \(0/0\) for \(\tau_i(t)\). Exact-flat participants are therefore not assigned a subjective-time trajectory for the primary analysis. Second, among raw non-flat participants, a numerical identification tolerance is applied after the normalized monotone reconstruction. The primary analysis requires
\begin{equation}
k_i\geq 10^{-8}.
\end{equation}
Raw non-flat participants with \(k_i<10^{-8}\) are classified as reconstruction-degenerate because the fitted discounting signal is numerically negligible relative to the normalization used to obtain \(\tau_i(t)\). The rule uses only the reconstructed discount curve and does not involve financial literacy or the other covariates; it does not, however, guarantee that the resulting selection is statistically unrelated to financial literacy, and the distribution of reconstruction-degenerate profiles across financial-literacy levels is reported in Table~\ref{tab:supp_identification_FL}. Sensitivity analyses repeat the primary model using identification tolerances \(0\), \(10^{-10}\), \(10^{-9}\), \(10^{-8}\), and \(10^{-7}\).

The centered functional response in Equation~\eqref{eq:tau_centered} is computed on the identified sample, with \(\bar{\tau}(t)\) the pointwise mean of the \(N=59\) identified trajectories.

\subsection{Flat Status, Near-Flat Profiles, and Complementary Full-Sample Response}

Exact-flat status is analyzed on the full analytical sample (\(N=139\)) with the logistic model
\begin{equation}
\begin{split}
\mathrm{logit}\,\Pr(\mathrm{flat}_i=1)={}&\gamma_0+\gamma_1FL_i\\
&+\gamma_2TSS_i+\gamma_3Age_i.
\end{split}
\end{equation}
This model addresses the occurrence of exact-flat raw response profiles and is distinct from the subsequent identification screening among raw non-flat participants.

To examine whether conclusions depend on small but nonzero raw discounting, a near-flat sensitivity analysis uses
\begin{equation}
1-\min_t f_i(t)\leq\epsilon,
\end{equation}
with \(\epsilon\in\{0,0.01,0.02,0.05,0.10\}\). The primary analysis corresponds to \(\epsilon=0\), which reproduces the exact-flat definition, together with the reconstruction-identification requirement \(k_i\geq10^{-8}\); positive values of \(\epsilon\) are reported as sensitivity analyses. The grid was fixed before inspecting the corresponding \(p\)-values.

A complementary full-sample analysis uses
\begin{equation}
g_i(t)=-\log\widehat f_i(t),
\end{equation}
which remains defined for exact-flat and reconstruction-degenerate participants because it does not divide by \(k_i\). The \(g_i(t)\) model therefore addresses the smoothed log-discount trajectory rather than the normalized subjective-time trajectory.

\subsection{Function-on-Scalar Regression}
\label{sec:methods_fosr}

The primary function-on-scalar model is
\begin{equation}
\label{eq:primary}
\begin{split}
\tau_i^c(t)={}&\beta_0(t)+\beta_1(t)FL_i+\beta_2(t)TSS_i\\
&+\beta_3(t)Age_i+\varepsilon_i(t),
\end{split}
\end{equation}
estimated on the \(N=59\) participants with an identified reconstructed trajectory. Financial literacy is coded numerically from \(0\) to \(3\) in the primary model. The coefficient function \(\beta_1(t)\) therefore represents the linear-score association between financial literacy and the centered reconstructed trajectory across the temporal domain. Four nested specifications are compared throughout and are labeled as follows: M1 contains financial literacy only; M2 contains financial literacy and \(\log(k_i)\); M3, the primary specification in Equation~\eqref{eq:primary}, contains financial literacy, TSS, and age; and M4 adds \(\log(k_i)\) to M3.

Each response trajectory is projected onto a cubic B-spline basis with \(15\) basis functions and a second-derivative roughness penalty with \(\lambda=100\); both values were fixed a priori. Sensitivity to the response representation and the monotone reconstruction is assessed through four alternative configurations, labeled S1 to S4 and compared with the primary configuration S0: \(10\) basis functions with \(\lambda=10\) (S1), \(20\) basis functions with \(\lambda=1000\) (S2), a coarser monotone reconstruction with basis dimensions \(8\) and \(6\) (S3), and a finer reconstruction with basis dimensions \(12\) and \(10\) (S4).

Global inference for financial literacy uses the conditional Freedman--Lane permutation procedure \citep{Freedman_Lane_1983}. The reduced model contains TSS and age, and the full model additionally contains financial literacy. The test statistic is the reduction in integrated residual sum of squares between the reduced and full functional models. Residual coefficient vectors from the reduced model are permuted across participants and added to the fitted reduced-model values before both models are re-estimated. The permutation distribution is based on \(B=10{,}000\) permutations, with the plus-one correction for the \(p\)-value. The partial functional coefficient of determination is
\begin{equation}
R^2_{\mathrm{partial}}=\frac{RSS_R-RSS_F}{RSS_R}.
\end{equation}

Uncertainty for \(\widehat\beta_{FL}(t)\) is assessed by a participant-level case bootstrap with \(B=1000\) resamples. Pointwise \(95\%\) bands use the 2.5th and 97.5th percentiles at each grid point. A simultaneous \(95\%\) band is additionally constructed with a bootstrap sup-\(t\) procedure, using the 95th percentile of the maximum absolute standardized bootstrap deviation over the domain.

Because the numerical \(0\)--\(3\) score imposes a linear ordering, an exploratory factor specification with four financial-literacy levels is also evaluated. A global factor-versus-controls test assesses whether financial-literacy level contributes to the functional response without imposing linearity, and a nested comparison of the factor specification with the numerical-score model tests departure from the linear-score representation. These analyses are interpreted as exploratory and are not used to replace the primary numerical-score model.

The M4 sensitivity model includes the reconstructed discounting scale:
\begin{equation}
\begin{split}
\tau_i^c(t)={}&\beta_0(t)+\beta_1(t)FL_i+\beta_2(t)TSS_i\\
&+\beta_3(t)Age_i+\beta_4(t)\log(k_i)+\varepsilon_i(t).
\end{split}
\end{equation}
Because \(k_i\) enters directly into the construction of \(\tau_i(t)\), \(\log(k_i)\) is not interpreted as an external independent confounder. This specification is used only to assess how strongly the estimated financial-literacy coefficient depends on conditioning on the scale component of the same reconstruction.

Further sensitivity analyses include educational attainment, exclusion of the \(\lceil0.04N\rceil=3\) trajectories with the largest value of \(\max_t|\tau_i''(t)|\), the near-flat grid, the reconstruction-identification tolerance grid, and an alternative penalized function-on-scalar estimator implemented in \texttt{refund::pffr}. The trimming fraction of \(4\%\) was fixed before inspecting the results. The relative \(L^2\) distance between coefficient functions is used to quantify changes in the estimated financial-literacy curve across specifications; a re-estimation excluding \(FL=0\) participants is reported on this scale only, as a coefficient-stability check rather than as an additional global test.

\subsection{Scalar Benchmark Models}

Scalar benchmark regressions are estimated on the same \(N=59\) identified sample. The outcomes are \(k_i\), \(\log(k_i)\), and the first two FPCA scores recomputed from the centered identified trajectories. Each outcome is regressed on financial literacy, TSS, and age. These models assess whether the numerical financial-literacy score is associated with scalar summaries of discount intensity or the dominant modes of functional variation, while the function-on-scalar model retains the full trajectory as the response \citep{Ramsay_2005}.

\section{Results}
\label{sec:results}

\subsection{Identification and Exact-Flat Status}

Among the \(139\) participants in the full analytical sample, \(61\) (\(43.9\%\)) had exact-flat raw discounting profiles and \(78\) were raw non-flat. All four financial-literacy levels were represented in both groups. Exact-flat status occurred in \(4\) of \(11\) participants with \(FL=0\), \(19\) of \(39\) with \(FL=1\), \(27\) of \(55\) with \(FL=2\), and \(11\) of \(34\) with \(FL=3\).

The logistic regression of exact-flat status on financial literacy, TSS, and age showed no evidence of association for any of the three predictors. For financial literacy, the odds ratio was \(0.860\) (95\% CI \(0.588\) to \(1.251\), \(p=0.430\)). The corresponding odds ratios were \(1.013\) for TSS (\(p=0.932\)) and \(0.998\) for age (\(p=0.897\)); McFadden's pseudo-\(R^2\) was \(0.0034\).

\begin{table}[htbp]
\centering
\caption{Logistic regression of exact-flat raw discounting status on financial literacy, TSS, and age (\(N=139\)). McFadden pseudo-\(R^2=0.0034\).}
\label{tab:logistic_flat}
\begin{tabular*}{\columnwidth}{@{\extracolsep{\fill}}lrrr@{}}
\toprule
Term & Odds ratio & 95\% CI & \(p\)-value \\
\midrule
FL  & 0.860 & [0.588, 1.251] & 0.430 \\
TSS & 1.013 & [0.754, 1.362] & 0.932 \\
Age & 0.998 & [0.970, 1.027] & 0.897 \\
\botrule
\end{tabular*}
\end{table}

The reconstruction step identified a second source of non-identification among the raw non-flat participants. Nineteen of the \(78\) raw non-flat profiles (\(24.4\%\)) produced a numerically reconstructed \(k_i<10^{-8}\) after the normalized monotone discount-curve fit. These profiles were classified as reconstruction-degenerate and were excluded from the primary \(\tau_i^c(t)\) analysis. The resulting primary identified sample contained \(N=59\) participants. By financial-literacy score, the identified sample contained \(5\) participants with \(FL=0\), \(11\) with \(FL=1\), \(23\) with \(FL=2\), and \(20\) with \(FL=3\). The corresponding numbers of reconstruction-degenerate raw non-flat participants were \(2\), \(9\), \(5\), and \(3\), respectively.

\subsection{Diagnostics and Scalar Benchmarks}

In the primary identified sample, mean financial literacy was \(1.98\) (SD \(0.94\)), mean TSS was \(4.91\) (SD \(1.13\)), and mean age was \(41.17\) years (SD \(13.29\)). Variance inflation factors were \(1.009\) for financial literacy, \(1.003\) for TSS, and \(1.013\) for age, indicating negligible collinearity among the predictors.

Financial literacy was weakly correlated with \(k_i\) (\(r=0.136\), \(p=0.306\)) and \(\log(k_i)\) (\(r=0.137\), \(p=0.301\)). The scalar benchmark regressions likewise did not provide evidence of a numerical financial-literacy association with \(k_i\) (\(p=0.355\)), \(\log(k_i)\) (\(p=0.308\)), FPC1 (\(p=0.260\)), or FPC2 (\(p=0.423\)). Thus, the scalar summaries do not indicate a clear monotonic association between financial literacy and discount intensity or the first two principal modes of functional variation in the identified sample.

\subsection{Descriptive Functional Profiles}

\begin{figure*}[htbp]
    \centering
    \includegraphics[width=\textwidth]{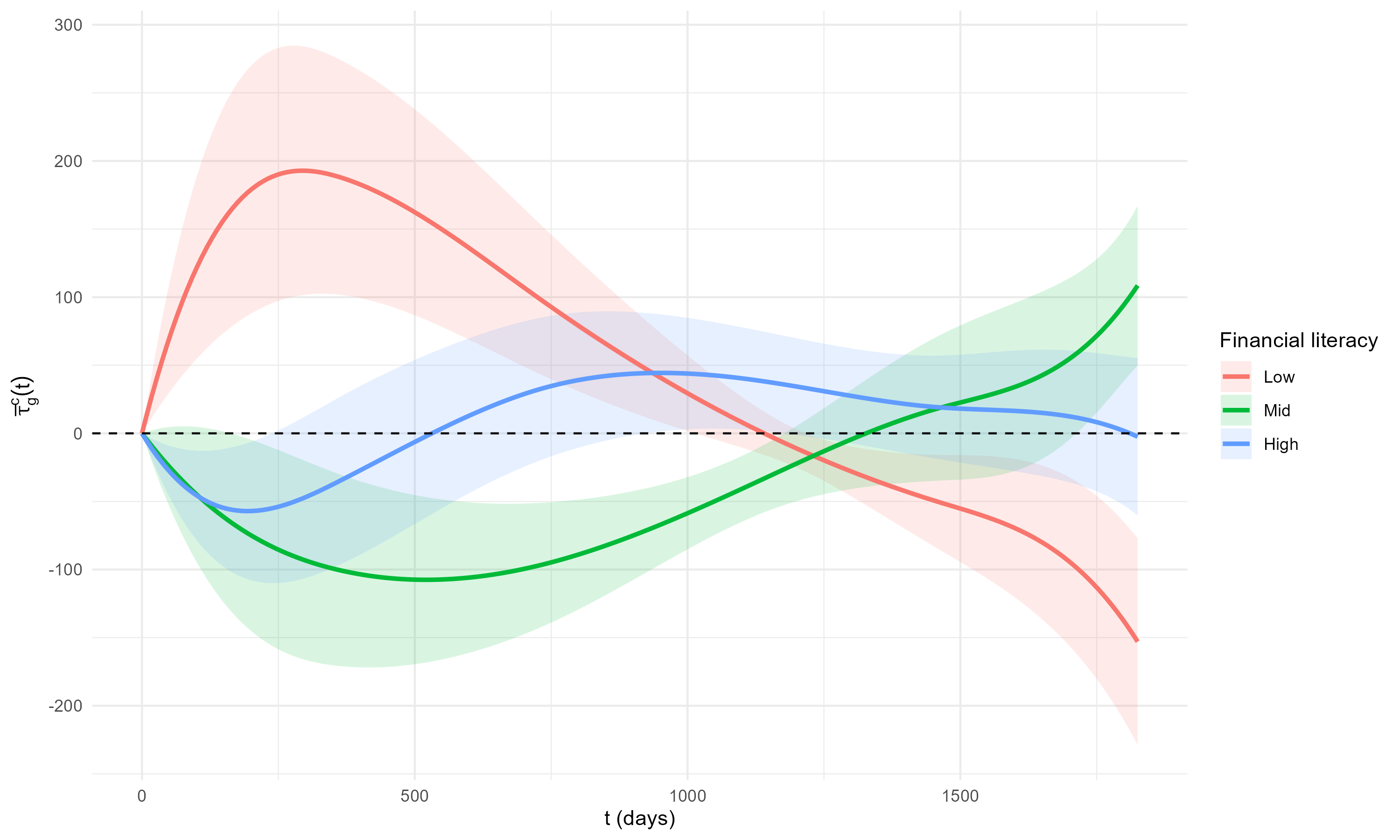}
    \caption{Mean centered reconstructed trajectories by financial-literacy group in the identified primary sample (low \(=FL\in\{0,1\}\), mid \(=FL=2\), high \(=FL=3\)). Shaded bands represent \(\pm1\) pointwise standard error and are descriptive.}
    \label{fig:fl_mean_profiles}
\end{figure*}

For descriptive visualization, the identified sample was grouped into low financial literacy (\(FL=0\) or \(1\), \(n=16\)), mid financial literacy (\(FL=2\), \(n=23\)), and high financial literacy (\(FL=3\), \(n=20\)). Figure~\ref{fig:fl_mean_profiles} shows the centered mean trajectories with pointwise standard-error bands. The group means do not follow a stable monotonic ordering across the domain and cross as the temporal horizon increases. This pattern suggests that a numerical linear score may not capture all between-level differences.

\subsection{Primary Function-on-Scalar Regression}

The primary function-on-scalar model treated financial literacy as a numerical \(0\)--\(3\) score and adjusted for TSS and age. The conditional Freedman--Lane test did not provide evidence that financial literacy improved the functional model relative to the controls-only specification (\(p=0.2578\), partial functional \(R^2=0.0216\)).

\begin{figure*}[htbp]
\centering
\includegraphics[width=\textwidth]{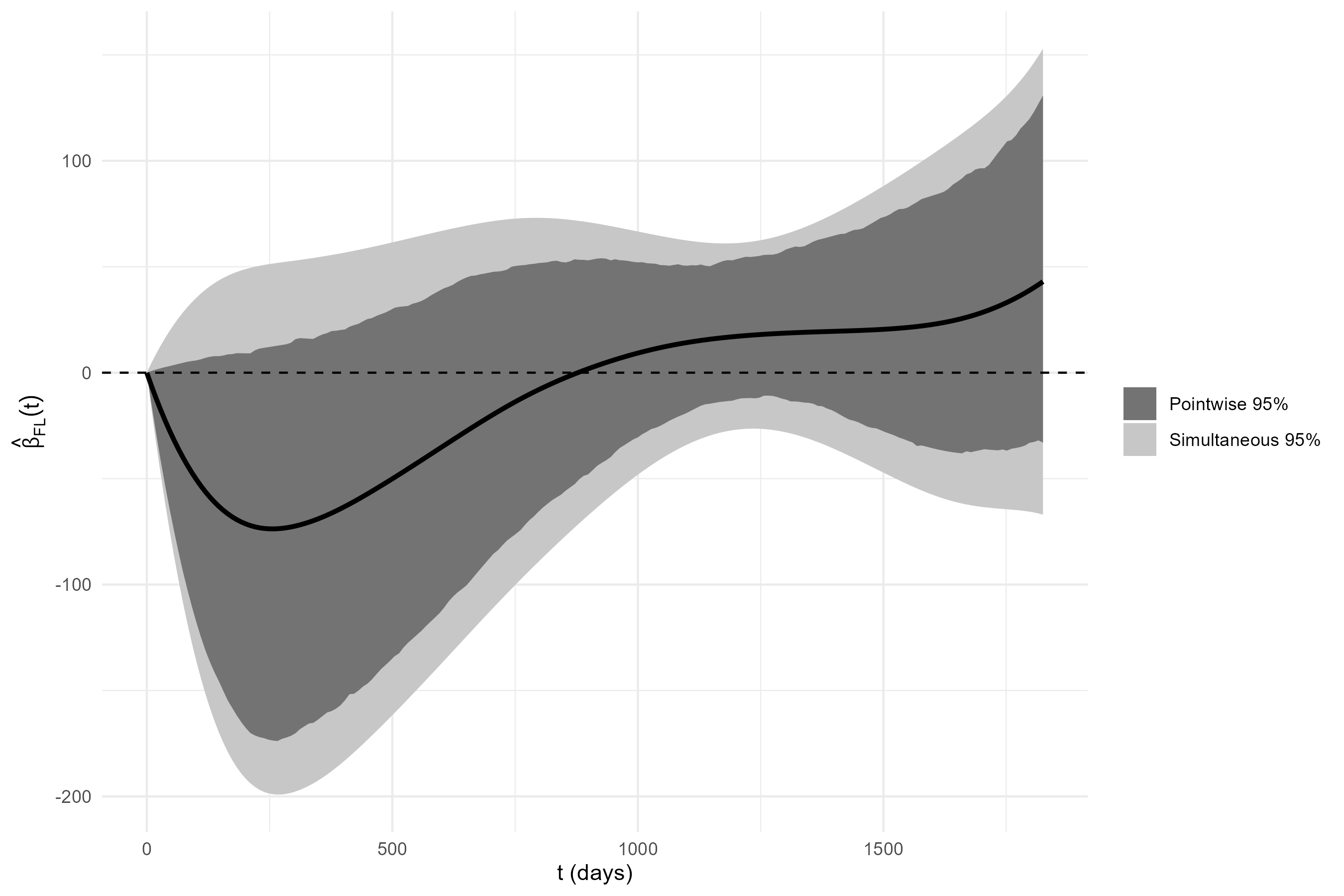}
\caption{Estimated financial-literacy coefficient function in the primary identified-sample model (\(N=59\)), with 95\% pointwise and simultaneous bootstrap bands. Neither band excludes zero at any point of the domain.}
\label{fig:beta_fl_simultaneous}
\end{figure*}

The pointwise bootstrap band covered zero over the entire observed domain. The simultaneous sup-\(t\) band also covered zero throughout, with a 95\% critical value of \(2.633\). Figure~\ref{fig:beta_fl_simultaneous} therefore shows no temporal interval in which the primary linear-score coefficient is distinguishable from zero under either band-based criterion.

The four-level categorical financial-literacy specification produced a different result. The global factor-versus-controls test was significant (\(p=0.0085\), partial functional \(R^2=0.1653\)), and the nested test comparing the factor model with the numerical-score model indicated departure from the linear-score specification (\(p=0.0049\), partial functional \(R^2=0.1468\)). These tests indicate heterogeneity across financial-literacy levels that a single linear score does not represent adequately in this sample. The individual level contrasts were, however, imprecise: the pointwise band excluded zero over \(27.5\%\) of the domain for \(FL=1\) versus \(FL=0\), and at no point of the domain for \(FL=2\) versus \(FL=0\) or \(FL=3\) versus \(FL=0\). Given the small \(FL=0\) reference group (\(n=5\)) and the exploratory status of the factor analysis, the clearest pairwise pattern involves \(FL=1\), and the results are interpreted as evidence of heterogeneity rather than as a stable level-specific ordering.

\subsection{Sensitivity Analyses}

The conclusion of no linear-score association was stable across the reconstruction-identification tolerance and near-flat analyses. The smallest reconstructed \(k_i\) in the identified sample was \(9.43\times10^{-7}\), so every positive tolerance from \(10^{-10}\) through \(10^{-7}\) selects the same \(N=59\) participants and returns the same test (\(p=0.258\), partial functional \(R^2=0.0216\)); the stability across this grid therefore reflects a gap in the distribution of \(k_i\) rather than independent replications. With no positive identification tolerance, all \(78\) raw non-flat participants entered the model, yielding \(p=0.690\) and partial functional \(R^2=0.0042\). Under raw near-flat exclusions, the primary \(\epsilon=0\) analysis yielded \(p=0.258\); the corresponding \(p\)-values were \(0.258\), \(0.134\), \(0.216\), and \(0.167\) for \(\epsilon=0.01\), \(0.02\), \(0.05\), and \(0.10\), respectively, where \(\epsilon=0.01\) excluded no additional participant and therefore reproduces the primary test.

The smoothing-sensitivity configurations also produced no significant linear financial-literacy test. The permutation \(p\)-values were \(0.263\) for S1, \(0.267\) for S2, \(0.438\) for S3, and \(0.259\) for S4. Excluding the three trajectories selected by the functional-outlier rule produced \(N=56\), \(p=0.340\), and partial functional \(R^2=0.0176\). Adding educational attainment yielded \(N=57\), \(p=0.352\), and partial functional \(R^2=0.0166\); educational attainment was recoded as an ordinal variable from middle school to doctorate, and the two identified participants who selected the residual ``Other'' category could not be placed on that scale and were excluded from this specification only.

The alternative penalized function-on-scalar estimator produced a coefficient curve highly correlated with the primary estimate (\(r=0.986\)); its relative \(L^2\) difference was \(0.163\), and its pointwise band also excluded zero nowhere on the domain. The absence of a linear-score association is therefore not specific to the basis-expansion estimator used in the primary analysis.

Re-estimating the primary model after excluding the five \(FL=0\) participants produced the largest change in the estimated coefficient function among all specifications considered (\(N=54\), relative \(L^2\) shift \(=1.274\)), consistent with the small size of that group.

Conditioning additionally on \(\log(k_i)\) changed the result. In this model, the global linear financial-literacy test yielded \(p=0.0465\) and partial functional \(R^2=0.0555\). However, the estimated financial-literacy coefficient changed substantially relative to the primary model, with a relative \(L^2\) shift of \(0.467\); its pointwise band excluded zero over \(38\%\) of the domain. Because \(\log(k_i)\) is mathematically coupled to the construction of \(\tau_i(t)\), and because the coefficient curve changes materially when it is included, this result is treated as specification-sensitive exploratory evidence rather than as confirmation of the primary model.

\begin{figure*}[htbp]
\centering
\includegraphics[width=\textwidth]{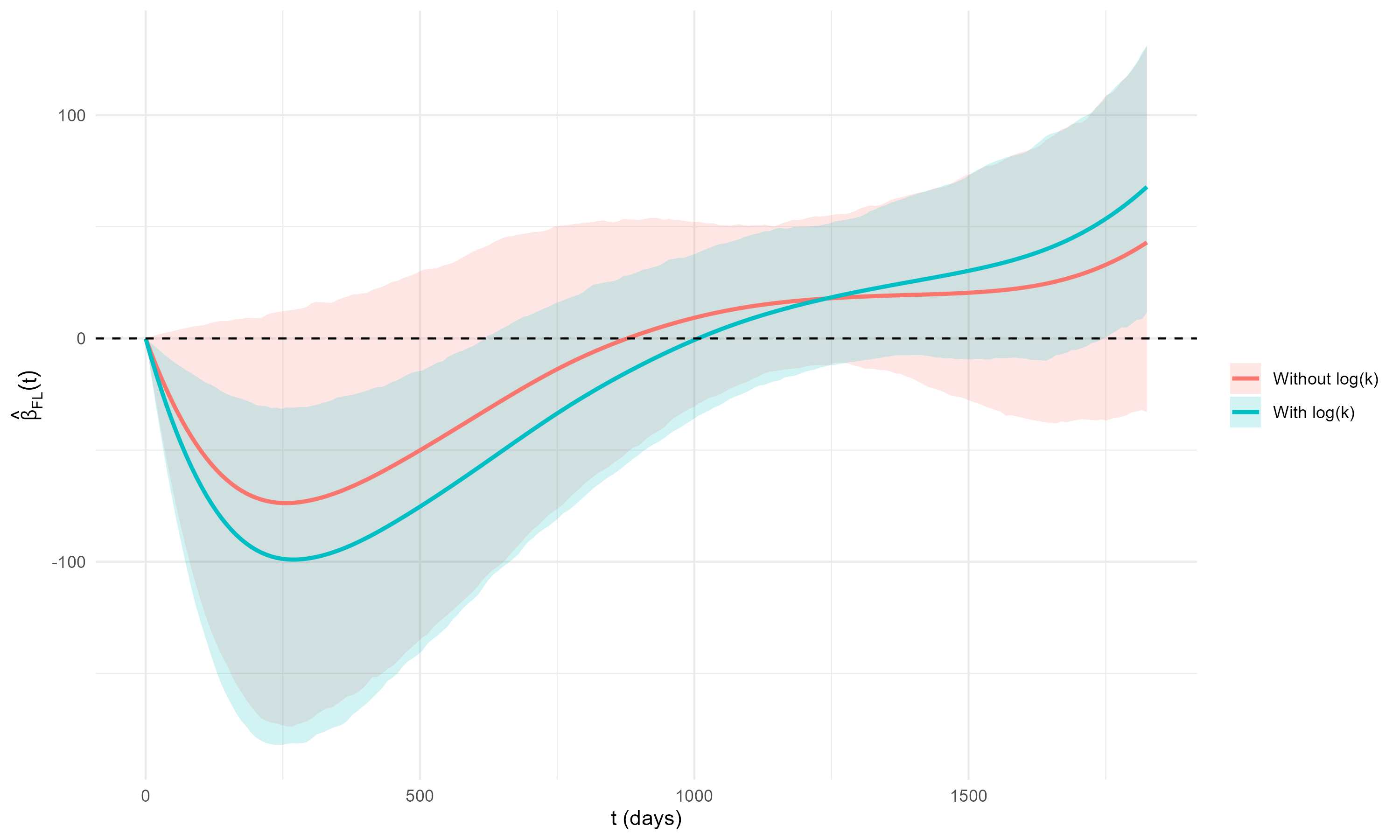}
\caption{Estimated financial-literacy coefficient function with and without \(\log(k_i)\) in the identified primary sample, with 95\% pointwise bootstrap bands.}
\label{fig:beta_fl_main_sensitivity}
\end{figure*}

\subsection{Complementary Full-Sample Analysis Using \(g(t)\)}

The complementary model based on \(g_i(t)=-\log\widehat f_i(t)\) was estimated on all \(139\) participants. The conditional Freedman--Lane test for the numerical financial-literacy score yielded \(p=0.1155\), with partial functional \(R^2=0.0166\). The corresponding four-level factor test was also not significant (\(p=0.1034\), partial functional \(R^2=0.0297\)).

For the numerical-score model, both the pointwise and the simultaneous \(95\%\) bands covered zero throughout the temporal domain; the simultaneous critical value was \(2.344\). Thus, the full-sample log-discount response does not provide evidence of a financial-literacy association under either the linear-score global test or the band-based analysis.

\begin{figure*}[htbp]
\centering
\includegraphics[width=\textwidth]{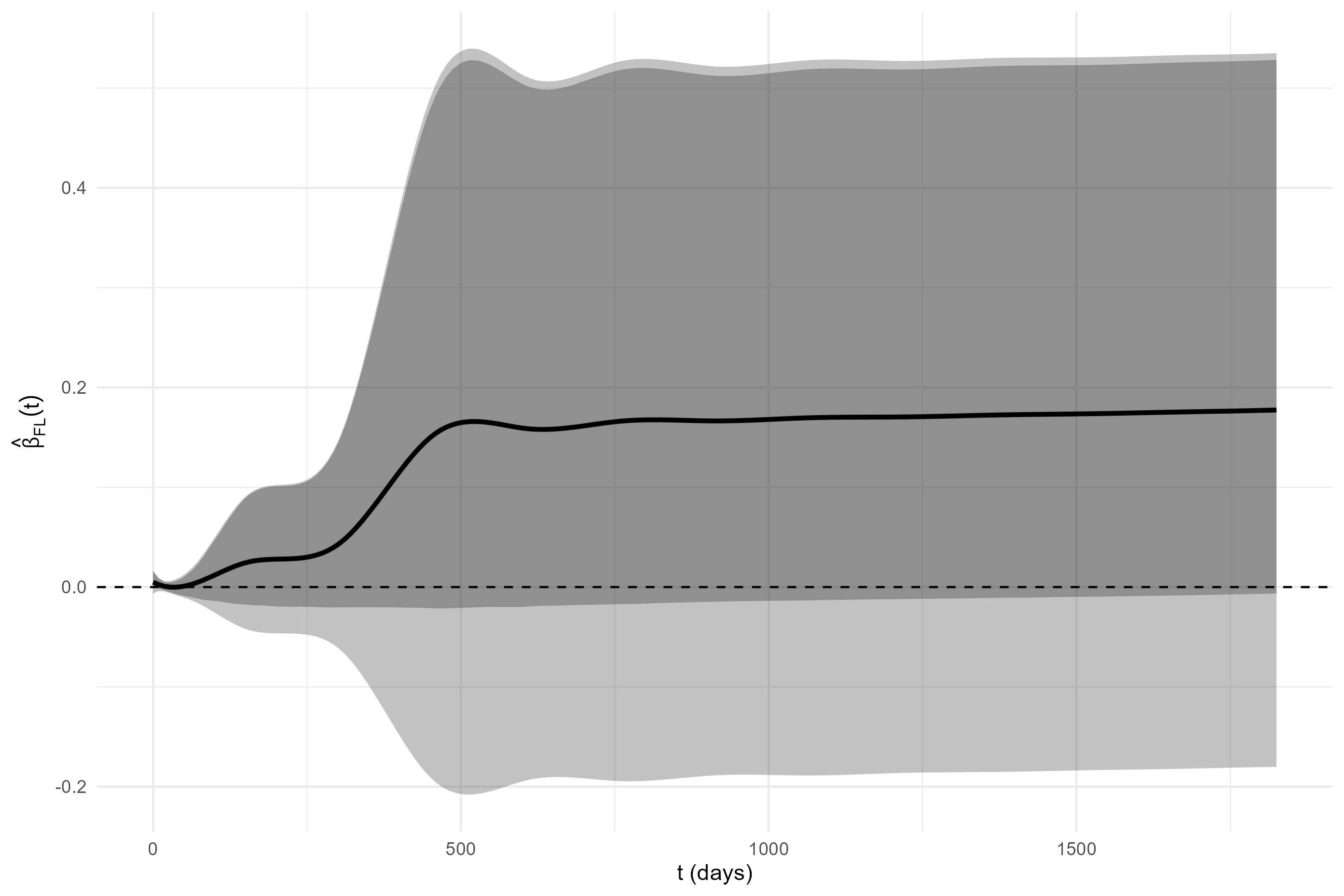}
\caption{Pointwise and simultaneous 95\% bands for the financial-literacy coefficient function in the complementary full-sample \(g(t)\) model (\(N=139\)).}
\label{fig:g_sup_t_band}
\end{figure*}

\section{Discussion}
\label{sec:discussion}

The primary analysis does not support a linear association between financial literacy and the centered reconstructed subjective-time trajectory in the subset for which that trajectory is identified. After excluding exact-flat profiles and raw non-flat profiles whose normalized monotone reconstruction produced \(k_i<10^{-8}\), the primary sample contained \(59\) participants. In this sample, the numerical financial-literacy score did not improve the function-on-scalar model after adjustment for TSS and age, and both pointwise and simultaneous uncertainty bands covered zero over the entire temporal domain. The scalar benchmark models similarly showed no clear association between the numerical financial-literacy score and \(k_i\), \(\log(k_i)\), FPC1, or FPC2.

The categorical analysis qualifies this result. When financial literacy was represented as a four-level factor, the global factor test was significant and the factor model provided a significantly better fit than the numerical-score model. The descriptive group trajectories also crossed over the domain rather than exhibiting a consistent low-to-high ordering. These results indicate that the absence of a numerical linear association should not be interpreted as evidence that all financial-literacy levels have identical functional profiles. The available data are compatible with heterogeneity across levels that a single \(0\)--\(3\) slope does not summarize. At the same time, the individual factor contrasts are imprecise, the clearest pairwise pattern involves \(FL=1\) against a reference group of only five participants, and the categorical analysis is exploratory. The factor result therefore motivates a hypothesis about departure from linearity rather than a stable ordering of financial-literacy levels.

The sensitivity analysis including \(\log(k_i)\) also requires a cautious interpretation. The global financial-literacy test becomes nominally significant when \(\log(k_i)\) is added, but the financial-literacy coefficient function changes substantially relative to the primary specification. Because \(k_i\) is estimated from the same reconstructed discount curve and enters directly into the normalization defining \(\tau_i(t)\), this sensitivity model does not provide an independent adjustment for an external covariate. The change in both inference and coefficient shape indicates that conclusions about financial literacy are sensitive to how the scale component of the reconstruction is handled, which argues against treating the \(\log(k_i)\)-adjusted result as a confirmation of a robust financial-literacy effect.

The complementary full-sample analysis based on \(g_i(t)=-\log\widehat f_i(t)\) likewise does not provide evidence of a numerical financial-literacy association. This result concerns a different functional response: \(g_i(t)\) represents the smoothed log-discount trajectory before normalization by \(k_i\), whereas \(\tau_i^c(t)\) represents deviations in the normalized reconstructed trajectory. The absence of evidence in the \(g(t)\) model therefore does not prove that financial literacy is unrelated to subjective temporal representation, but it limits the extent to which the exploratory categorical pattern observed for \(\tau_i^c(t)\) can be generalized to the full sample or to the discount trajectory itself.

Exact-flat status and reconstruction-level identification must be kept distinct. Exact-flat status is a property of the raw equivalence judgments and was not associated with financial literacy, TSS, or age in the logistic model. Reconstruction degeneracy is different: it occurs among raw non-flat participants when the monotone fitted discount curve contains too little signal to support stable division by \(k_i\). The primary estimand is therefore conditional on having an identified normalized trajectory after reconstruction, a conditioning that applies to \(19\) of the \(78\) raw non-flat participants.

Functional regression can retain information on the configuration of the entire trajectory and can test whether covariate associations vary over the temporal domain. However, when the response itself is obtained through a ratio involving an estimated scale parameter, identification of that ratio must be addressed before inferential interpretation. In the present data, the conclusion of no linear association holds both on the identified sample and on all \(78\) raw non-flat participants, while the analytical sample and the estimated coefficient function differ appreciably between the two.

The study is observational and exploratory. None of the reported associations can be interpreted causally, and financial literacy should not be interpreted as a proxy for rationality or decision quality. The principal empirical finding is that there is no robust evidence of a linear financial-literacy association with the identified reconstructed trajectory, while an exploratory factor specification indicates departure from linearity that should be evaluated in larger, independent samples.

\section{Limitations}
\label{sec:limitations}

Several limitations affect the interpretation and generalizability of the results. First, the sample was obtained through convenience and snowball recruitment and was not designed to represent a defined population. The full analytical sample contains \(139\) participants, but the primary functional analysis contains only \(59\). The primary result therefore pertains specifically to participants for whom the normalized reconstructed trajectory is identified under the adopted procedure, not to the full sample of respondents.

Second, the reconstruction of \(\tau_i(t)\) depends on monotone smoothing, the estimated discounting scale \(k_i\), and the numerical identification tolerance. The smoothing alternatives did not change the primary inferential conclusion, but these checks do not eliminate reconstruction uncertainty. In particular, uncertainty in the estimation of \(k_i\) is not propagated jointly through every stage of the subsequent functional regression. The marked change in the financial-literacy coefficient after inclusion of \(\log(k_i)\) also shows that the fitted association is sensitive to how the scale component of the reconstruction is represented. No correction for multiplicity was applied across the sensitivity analyses reported in Section~\ref{sec:results} and the Appendix, so the nominal significance of the \(\log(k_i)\)-adjusted test should be read in light of the number of specifications examined.

Third, the categorical financial-literacy result is exploratory and should not be overinterpreted. The identified sample contains only \(5\) participants with \(FL=0\), and the factor analysis estimates several functional contrasts. Although the global factor and departure-from-linearity tests are significant, the individual level contrasts are much less precise, and the coefficient function is highly sensitive to the exclusion of the \(FL=0\) group. The data therefore do not support a stable threshold, ordering, or specific non-linear functional form.

Fourth, the logistic model concerns exact-flat raw status only. It does not establish that the \(59\)-participant identified sample is a random subset of the \(78\) raw non-flat participants. Reconstruction-degenerate profiles are descriptively distributed differently across financial-literacy levels, and no separate inferential selection model was specified for that status. Selection into the identified functional sample therefore remains a potential limitation.

Fifth, local features of the reconstructed trajectories are less stable near the boundaries of the temporal domain, and derivative-based quantities are additionally affected by the smoothing step. Local interpretation close to the boundaries should therefore be made cautiously. The permutation procedure assumes exchangeability of the reduced-model residuals, and the within-group functional variability of the response differs across financial-literacy groups, so the nominal level of the test may be affected.

Finally, the questionnaire was administered in Italian, English, and Spanish, but the sample is overwhelmingly Italian-speaking. The language versions were not subjected to formal back-translation or independent linguistic validation, and the non-Italian subsamples are too small for separate analysis. The intertemporal equivalence task was hypothetical and non-incentivized, so the extent to which the observed patterns would replicate under incentivized elicitation cannot be determined. Larger and more heterogeneous samples are needed both to improve precision at the individual financial-literacy levels and to evaluate the identification characteristics of the reconstruction in other populations.

\section{Conclusions}
\label{sec:conclusions}

The analysis does not provide robust evidence of a linear association between financial literacy and the centered reconstructed subjective-time trajectory among participants for whom that trajectory is identified. In the primary \(N=59\) function-on-scalar model, the numerical financial-literacy score was not globally significant and both pointwise and simultaneous confidence bands covered zero throughout the domain. The complementary full-sample \(g(t)\) analysis reached the same broad conclusion for the numerical score.

An exploratory categorical specification nevertheless indicated significant differences among financial-literacy levels and significant departure from the linear-score representation. Because the identified sample is small at some financial-literacy levels and the individual contrasts are imprecise, this pattern should be regarded as a hypothesis about departure from linearity rather than as a definitive functional ordering. The sensitivity of the coefficient function to inclusion of \(\log(k_i)\) further reinforces the need for caution.

The methodological contribution of the present study is the explicit separation of raw exact-flat profiles from reconstruction-level non-identification when a reconstructed trajectory is used as a functional response. Future work should evaluate the categorical pattern in larger independent samples, develop reconstruction and inferential procedures that propagate uncertainty in \(k_i\), and examine whether alternative elicitation designs reduce the frequency of exact-flat and reconstruction-degenerate profiles.

\backmatter

\bmhead{Supplementary information}

The R code used to reproduce the analyses is provided as Online Resource.

\bmhead{Acknowledgements}

The authors thank the participants who completed the questionnaire.

\section*{Statements and Declarations}

\begin{itemize}
\item \textbf{Funding.} The authors declare that no funds, grants, or other
support were received during the preparation of this manuscript.

\item \textbf{Competing interests.} The authors have no relevant financial
or non-financial interests to disclose.

\item \textbf{Ethics approval.} The study was conducted in accordance with
the principles of the Declaration of Helsinki. It involved an anonymous,
non-clinical, minimal-risk online questionnaire administered to adult
participants, with no clinical or invasive procedures, biological samples,
deception, or vulnerable populations. The study did not fall within the
categories of research subject to mandatory institutional ethics review at
the participating institutions, and no formal ethics approval number was
issued.

\item \textbf{Consent to participate.} Participation was voluntary and
uncompensated. Before beginning the questionnaire, participants were
informed of the scientific purpose of the study, the voluntary nature of
participation, their right to discontinue participation at any time, and the
intended anonymous and aggregate use of their responses for scientific
research. No directly identifying information, including names, contact
details, or IP addresses, was collected or retained. Participants provided
informed consent by voluntarily proceeding with the questionnaire after
receiving this information. Data processing was conducted in accordance with
Regulation (EU) 2016/679 (GDPR) and applicable Italian data protection
requirements.

\item \textbf{Consent for publication.} All authors consent to the
publication of this manuscript.

\item \textbf{Data availability.} The data supporting the findings of this
study were collected through an anonymous online questionnaire and are not
publicly available because of privacy and confidentiality considerations. An
anonymized version of the dataset (\(N=140\) participants, \(139\) with a
valid discount-curve reconstruction) may be made available by the corresponding
author upon reasonable request, subject to applicable data protection
requirements.

\item \textbf{Code availability.} The R code used to reproduce the analyses
is provided as Online Resource.

\item \textbf{Author contributions.} All authors contributed equally to the study
conception and design, data collection, statistical analysis, interpretation of
the results, and the writing and critical revision of the manuscript. All
authors read and approved the final manuscript.
\end{itemize}

\clearpage
\onecolumn
\begin{appendices}
\counterwithin*{table}{section}
\counterwithin*{figure}{section}
\renewcommand{\theHtable}{app.\arabic{table}}
\renewcommand{\theHfigure}{app.\arabic{figure}}
\section{Sample flow and identification}\label{secA1}

This appendix reports the identification diagnostics, scalar benchmarks, functional tests, and sensitivity analyses referred to in the main text.

The raw dataset contained \(140\) participants. One participant had only \(4\) valid equivalence observations and did not meet the requirement of at least \(6\) valid elicited horizons for discount-curve smoothing. The full analytical sample therefore contains \(139\) participants. Exact-flat status is defined directly from the raw equivalence judgments. Reconstruction-level identification is assessed separately among raw non-flat participants after monotone smoothing and normalization of the fitted discount curve to \(\widehat f_i(0)=1\). The sample flow is reported in Table~\ref{tab:sample_description} of the main text.

\begin{blocktable}
\captionof{table}{Exact-flat and raw non-flat status by financial-literacy score in the full analytical sample.}
\label{tab:supp_flat_crosstab}
\begin{tabular}{@{}lrrr@{}}
\toprule
Financial literacy & Raw non-flat & Exact-flat & Total \\
\midrule
0 & 7 & 4 & 11 \\
1 & 20 & 19 & 39 \\
2 & 28 & 27 & 55 \\
3 & 23 & 11 & 34 \\
\midrule
Total & 78 & 61 & 139 \\
\botrule
\end{tabular}
\end{blocktable}

\begin{blocktable}
\captionof{table}{Reconstruction-identification status by financial-literacy score.}
\label{tab:supp_identification_FL}
\begin{tabular}{@{}lrrr@{}}
\toprule
Financial literacy & Exact-flat & Reconstruction-degenerate & Identified \(\tau_i(t)\) \\
\midrule
0 & 4 & 2 & 5 \\
1 & 19 & 9 & 11 \\
2 & 27 & 5 & 23 \\
3 & 11 & 3 & 20 \\
\midrule
Total & 61 & 19 & 59 \\
\botrule
\end{tabular}
\footnotetext{The reconstruction-degenerate column refers to raw non-flat participants only.}
\end{blocktable}

Figure~\ref{fig:supp_ki_identification} shows the distribution of the reconstructed discount-intensity parameter among the \(78\) raw non-flat participants, distinguishing profiles that satisfy the primary identification requirement from reconstruction-degenerate profiles. Exact-flat participants are not shown because \(k_i=0\) for them by construction.

\begin{blockfigure}
\includegraphics[width=0.85\textwidth]{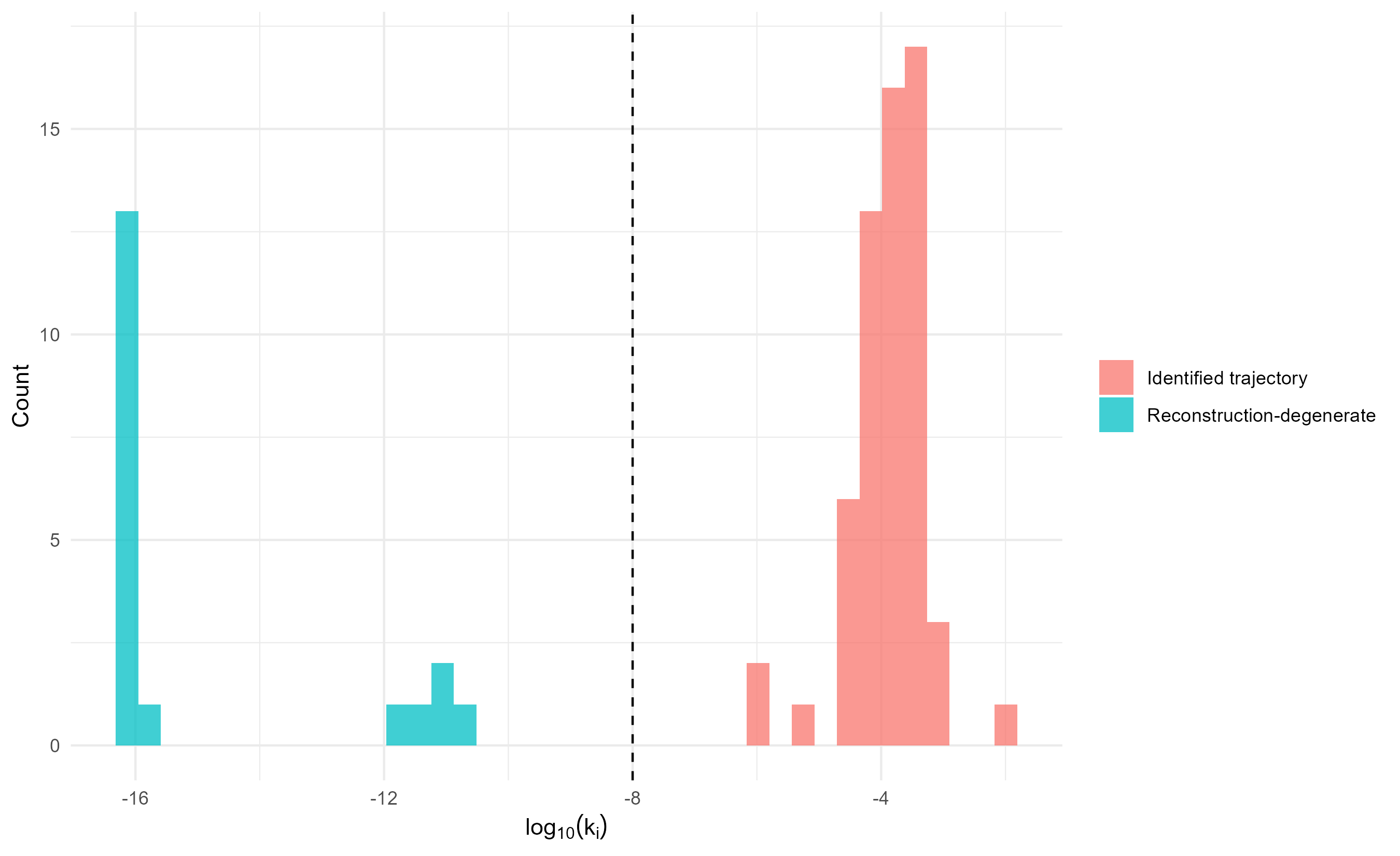}
\captionof{figure}{Distribution of the reconstructed discounting scale \(k_i\), on the \(\log_{10}\) scale, among the \(78\) raw non-flat participants, by identification status. The vertical reference is the identification tolerance \(10^{-8}\). No reconstructed value lies between \(2.03\times10^{-11}\) and \(9.43\times10^{-7}\).}
\label{fig:supp_ki_identification}
\end{blockfigure}

\clearpage
\section{Primary identified-sample analysis}\label{secA2}

This appendix reports descriptive statistics, scalar diagnostics, global tests, coefficient functions, and sensitivity analyses for the primary identified sample (\(N=59\)).

\FloatBarrier
\Needspace{0.45\textheight}
\subsection{Descriptive statistics and scalar diagnostics}

\begin{blocktable}
\captionof{table}{Descriptive statistics of the primary identified sample (\(N=59\)).}
\label{tab:supp_desc_identified}
\begin{tabular}{@{}lrrrrrr@{}}
\toprule
Variable & \(n\) & Mean & SD & Min & Median & Max \\
\midrule
FL  & 59 & 1.98 & 0.94 & 0 & 2 & 3 \\
TSS & 59 & 4.91 & 1.13 & 2.2 & 5 & 7 \\
Age & 59 & 41.17 & 13.29 & 18 & 41 & 73 \\
\botrule
\end{tabular}
\end{blocktable}

\begin{blocktable}
\captionof{table}{Pearson correlation matrix in the primary identified sample.}
\label{tab:supp_corr_identified}
\small
\begin{tabular}{@{}lrrrrrrr@{}}
\toprule
 & FL & TSS & Age & \(k_i\) & \(\log(k_i)\) & FPC1 & FPC2 \\
\midrule
FL & 1.000 & -0.011 & -0.097 & 0.136 & 0.137 & -0.146 & -0.123 \\
TSS & -0.011 & 1.000 & 0.059 & -0.092 & 0.074 & -0.048 & 0.124 \\
Age & -0.097 & 0.059 & 1.000 & -0.128 & -0.011 & -0.045 & 0.172 \\
\(k_i\) & 0.136 & -0.092 & -0.128 & 1.000 & 0.486 & 0.124 & -0.300 \\
\(\log(k_i)\) & 0.137 & 0.074 & -0.011 & 0.486 & 1.000 & 0.513 & -0.229 \\
FPC1 & -0.146 & -0.048 & -0.045 & 0.124 & 0.513 & 1.000 & 0.009 \\
FPC2 & -0.123 & 0.124 & 0.172 & -0.300 & -0.229 & 0.009 & 1.000 \\
\botrule
\end{tabular}
\end{blocktable}

\begin{blocktable}
\captionof{table}{Variance inflation factors for the predictors in the primary functional model.}
\label{tab:supp_vif_identified}
\begin{tabular*}{0.5\textwidth}{@{\extracolsep{\fill}}lr@{}}
\toprule
Predictor & VIF \\
\midrule
FL & 1.009 \\
TSS & 1.003 \\
Age & 1.013 \\
\botrule
\end{tabular*}
\end{blocktable}

\begin{blocktable}
\captionof{table}{Financial-literacy coefficient in scalar benchmark regressions (\(N=59\)).}
\label{tab:supp_scalar_benchmarks}
\begin{tabular}{@{}lrrrr@{}}
\toprule
Outcome & Estimate & SE & \(p\)-value & Model \(R^2\) \\
\midrule
\(k_i\) & \(1.73\times10^{-4}\) & \(1.86\times10^{-4}\) & 0.355 & 0.0387 \\
\(\log(k_i)\) & 0.220 & 0.214 & 0.308 & 0.0245 \\
FPC1 & -1463.69 & 1286.90 & 0.260 & 0.0269 \\
FPC2 & -404.22 & 501.13 & 0.423 & 0.0537 \\
\botrule
\end{tabular}
\end{blocktable}

\FloatBarrier
\Needspace{0.45\textheight}
\subsection{Primary function-on-scalar tests}

\begin{blocktable}
\captionof{table}{Conditional Freedman--Lane global tests for financial literacy in the identified sample (\(B=10{,}000\)).}
\label{tab:supp_formal_tests}
\begin{tabular}{@{}lrrr@{}}
\toprule
Test & \(N\) & \(p\)-value & Partial functional \(R^2\) \\
\midrule
Numerical FL score \(\mid\) TSS + Age & 59 & 0.2578 & 0.0216 \\
FL factor \(\mid\) TSS + Age & 59 & 0.0085 & 0.1653 \\
Departure from numerical score & 59 & 0.0049 & 0.1468 \\
\botrule
\end{tabular}
\end{blocktable}

\begin{blockfigure}
\includegraphics[width=\textwidth]{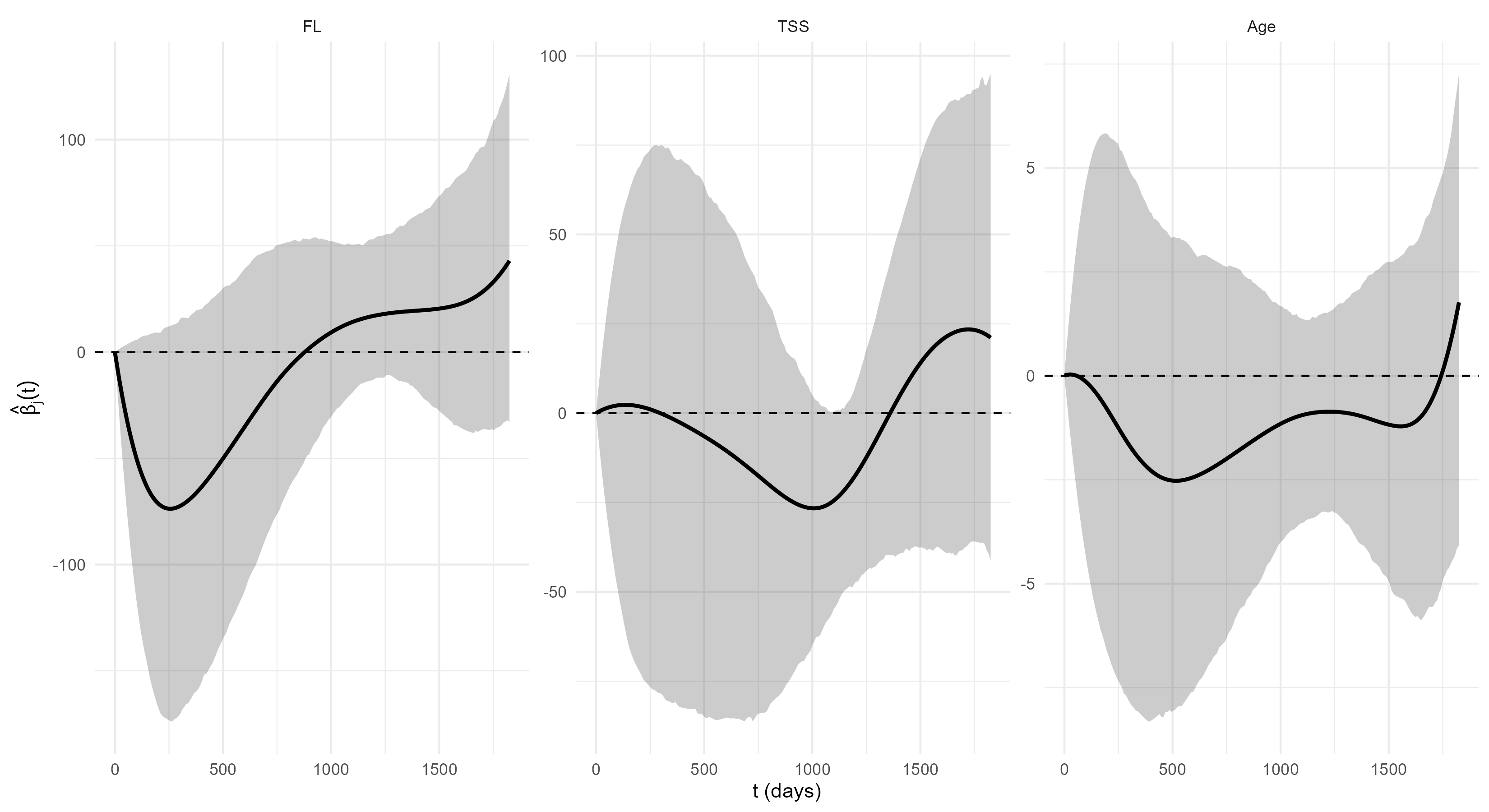}
\captionof{figure}{Coefficient functions for financial literacy, the Temporal Sense Scale, and age in the primary identified-sample model (\(N=59\)), with 95\% pointwise bootstrap bands.}
\label{fig:supp_beta_primary}
\end{blockfigure}

\begin{blocktable}
\captionof{table}{\(L^2\) norm of the financial-literacy coefficient across the four nested specifications defined in Section~\ref{sec:methods_fosr}.}
\label{tab:supp_L2_models}
\begin{tabular*}{0.5\textwidth}{@{\extracolsep{\fill}}lr@{}}
\toprule
Specification & \(L^2(\widehat\beta_{FL})\) \\
\midrule
M1: FL & 1551.06 \\
M2: FL + \(\log(k_i)\) & 2208.07 \\
M3: FL + TSS + Age (primary) & 1592.94 \\
M4: FL + TSS + Age + \(\log(k_i)\) & 2270.35 \\
\botrule
\end{tabular*}
\end{blocktable}

\begin{blocktable}
\captionof{table}{Pointwise and simultaneous band summary for the financial-literacy coefficient in the primary specification.}
\label{tab:supp_band_primary}
\begin{tabular}{@{}lr@{}}
\toprule
Quantity & Value \\
\midrule
Simultaneous 95\% critical value & 2.633 \\
Share of domain with pointwise band excluding zero & 0.000 \\
Share of domain with simultaneous band excluding zero & 0.000 \\
\botrule
\end{tabular}
\end{blocktable}

\FloatBarrier
\Needspace{0.45\textheight}
\subsection{Alternative estimator}

\begin{blocktable}
\captionof{table}{Agreement between the primary estimator and \texttt{refund::pffr}.}
\label{tab:supp_pffr}
\begin{tabular}{@{}lr@{}}
\toprule
Metric & Value \\
\midrule
Correlation between coefficient curves & 0.986 \\
Relative \(L^2\) difference & 0.163 \\
Maximum absolute difference & 33.323 \\
Share of domain with \texttt{pffr} pointwise band excluding zero & 0.000 \\
\botrule
\end{tabular}
\end{blocktable}

\begin{blockfigure}
\includegraphics[width=\textwidth]{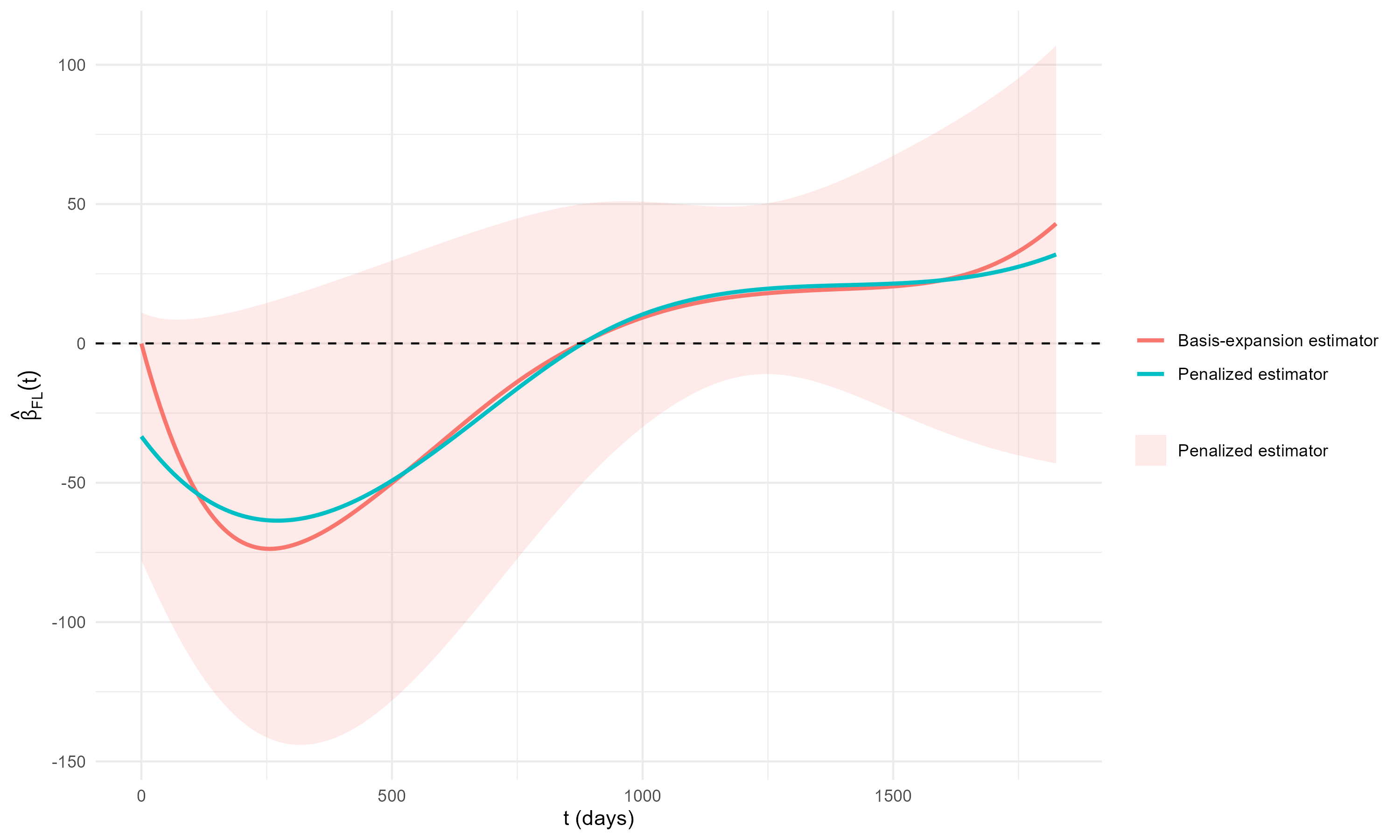}
\captionof{figure}{Comparison of the primary financial-literacy coefficient with the penalized \texttt{refund::pffr} estimate (\(r=0.986\), relative \(L^2\) difference \(=0.163\)). The shaded band is the 95\% pointwise interval of the penalized estimate.}
\label{fig:supp_pffr}
\end{blockfigure}

\FloatBarrier
\Needspace{0.45\textheight}
\subsection{Categorical and grouped financial-literacy contrasts}

The identified sample contains \(5\), \(11\), \(23\), and \(20\) participants at \(FL=0,1,2,3\), respectively. The grouped representation contains \(16\) low-\(FL\) participants (\(FL=0\) or \(1\)), \(23\) mid-\(FL\) participants (\(FL=2\)), and \(20\) high-\(FL\) participants (\(FL=3\)).

\begin{blockfigure}
\includegraphics[width=\textwidth]{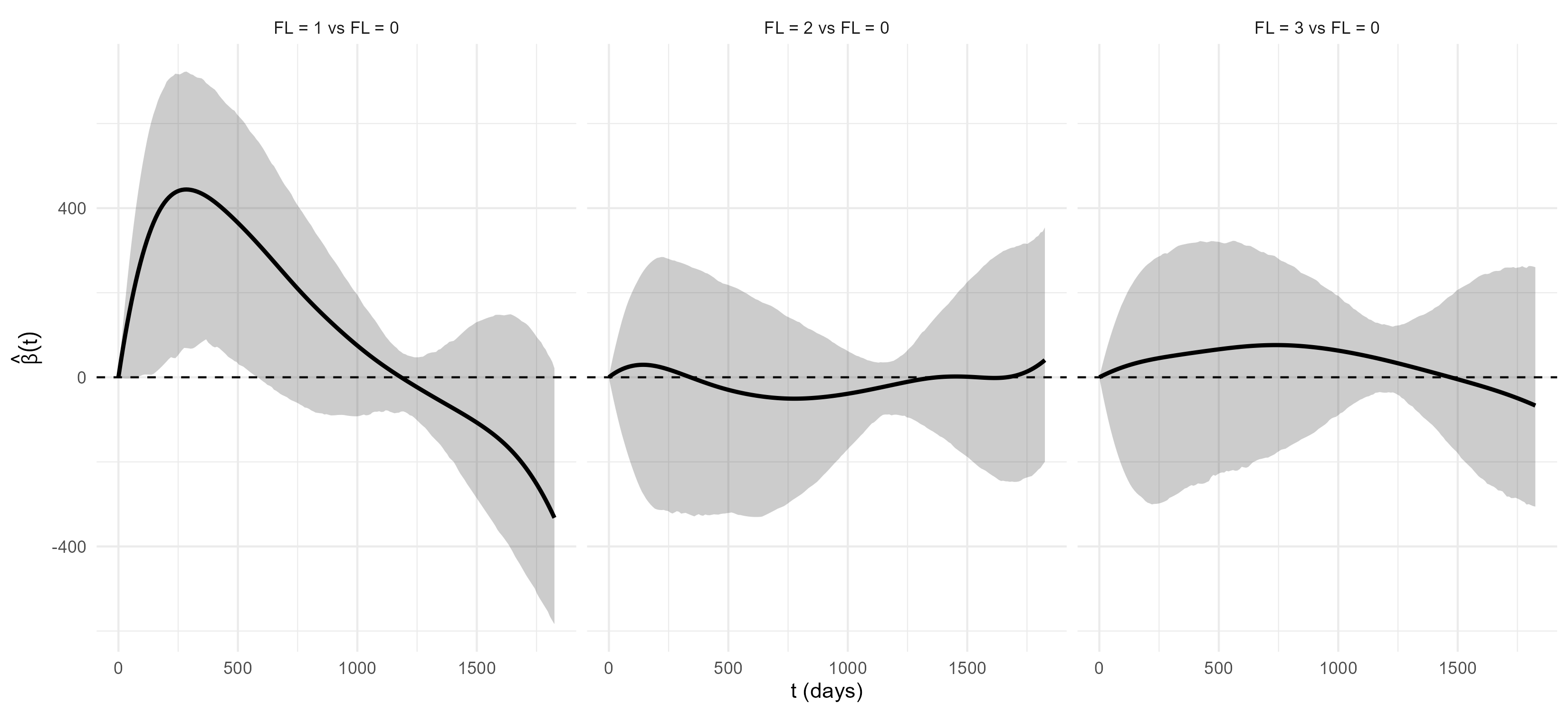}
\captionof{figure}{Categorical financial-literacy contrasts in the identified sample, with \(FL=0\) as reference. The 95\% pointwise band excludes zero over \(27.5\%\) of the domain for \(FL=1\) versus \(FL=0\), and at no point of the domain for \(FL=2\) versus \(FL=0\) and \(FL=3\) versus \(FL=0\).}
\label{fig:supp_categorical}
\end{blockfigure}

\begin{blockfigure}
\includegraphics[width=\textwidth]{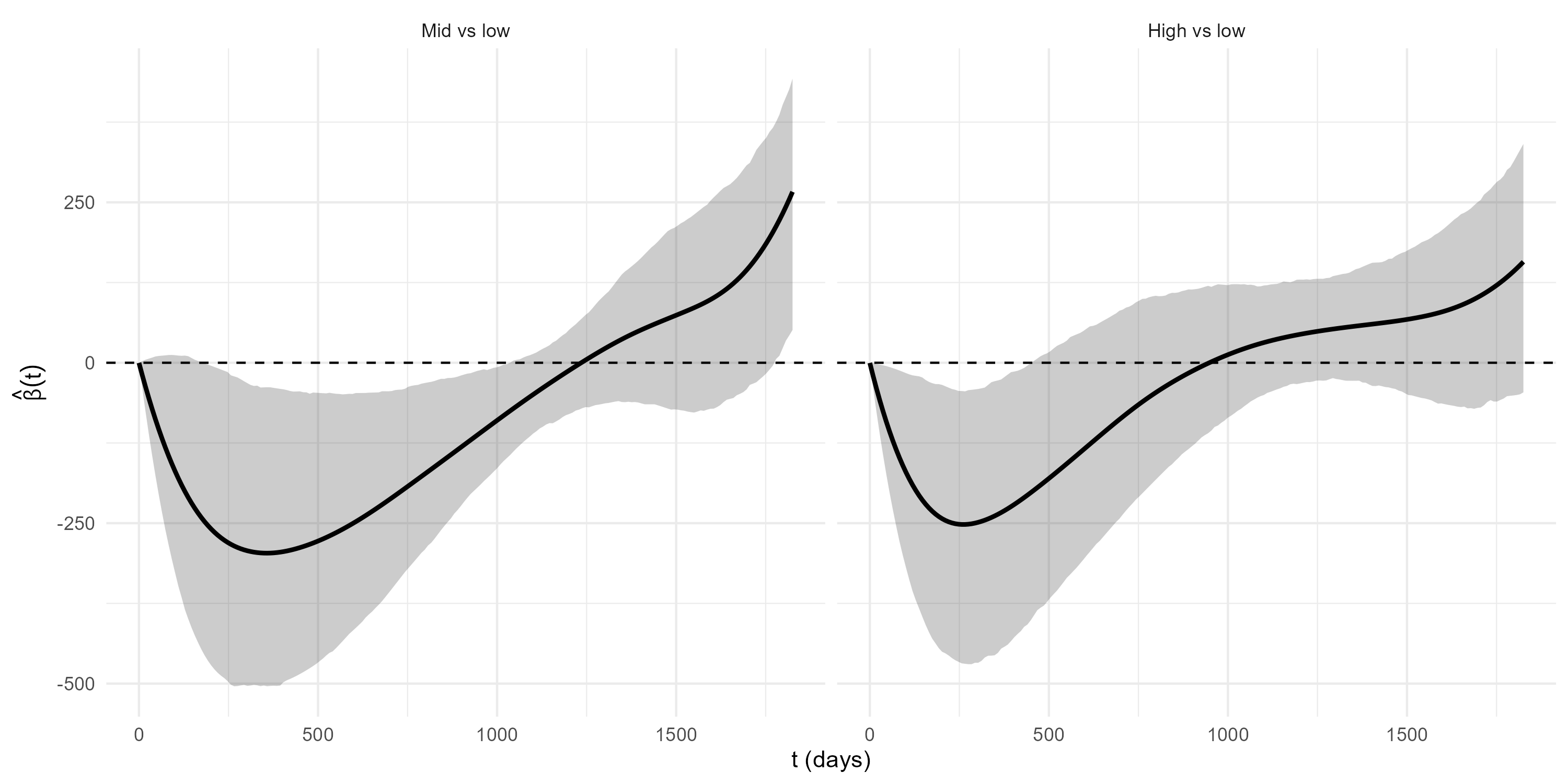}
\captionof{figure}{Grouped financial-literacy contrasts in the identified sample, with the low group as reference. The 95\% pointwise band excludes zero over \(49.5\%\) of the domain for mid versus low and over \(24.5\%\) for high versus low. These pointwise comparisons are descriptive and are not simultaneous tests.}
\label{fig:supp_grouped}
\end{blockfigure}

\begin{blockfigure}
\includegraphics[width=\textwidth]{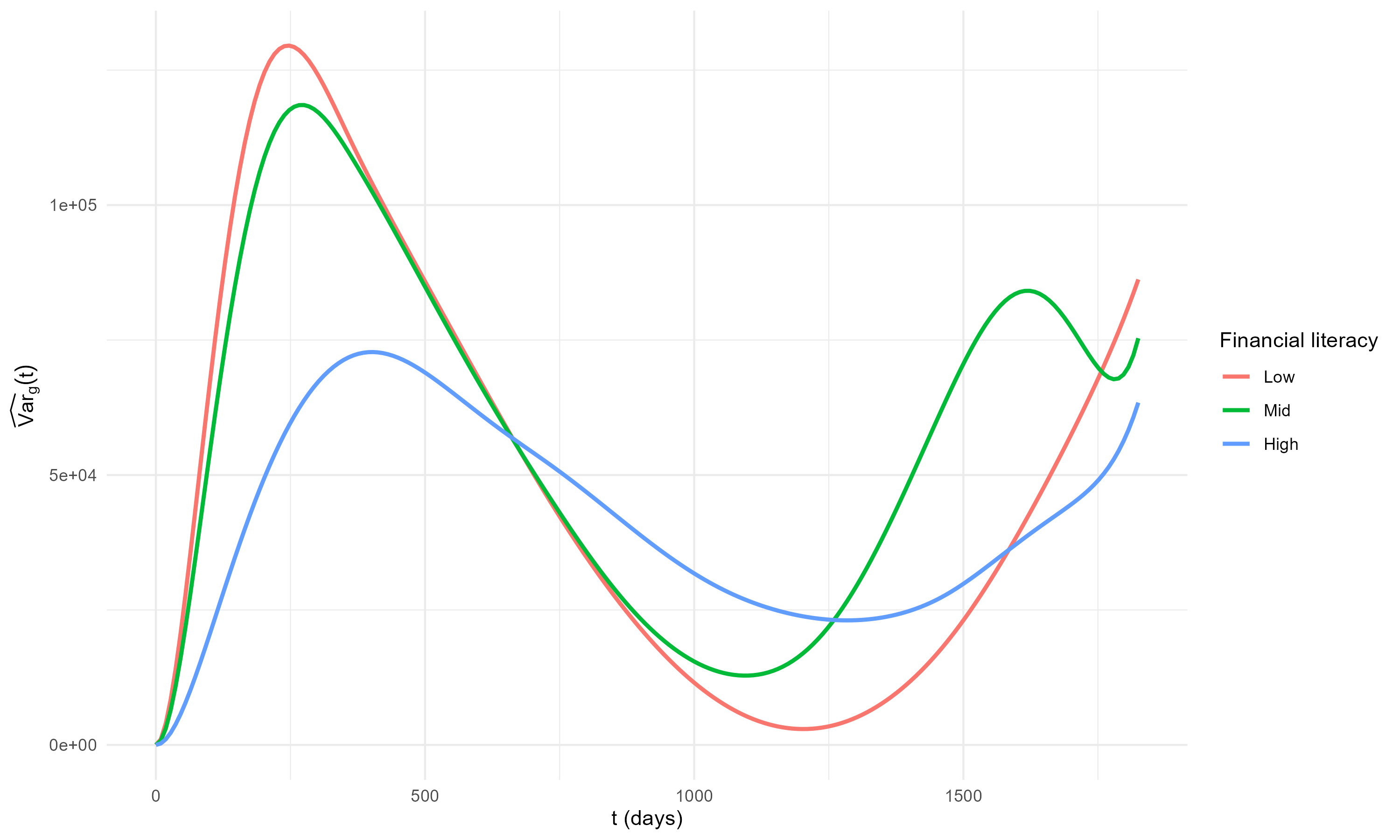}
\captionof{figure}{Pointwise functional variance of the centered reconstructed trajectories by grouped financial-literacy level.}
\label{fig:supp_group_variance}
\end{blockfigure}

\FloatBarrier
\Needspace{0.45\textheight}
\subsection{Sensitivity analyses}

\begin{blocktable}
\captionof{table}{Relative \(L^2\) changes in the estimated financial-literacy coefficient, relative to the primary specification M3.}
\label{tab:supp_L2_shifts}
\begin{tabular}{@{}lrr@{}}
\toprule
Comparison & \(N\) & Relative \(L^2\) shift \\
\midrule
M1 vs. M3 & 59 & 0.057 \\
M2 vs. M3 & 59 & 0.418 \\
M4 vs. M3 & 59 & 0.467 \\
Excluding \(FL=0\) vs. M3 & 54 & 1.274 \\
Outlier-trimmed vs. M3 & 56 & 0.143 \\
\texttt{refund::pffr} vs. primary estimator & 59 & 0.163 \\
Education-adjusted vs. M3 & 57 & 0.178 \\
\botrule
\end{tabular}
\end{blocktable}

\begin{blocktable}
\captionof{table}{Sensitivity analyses for educational attainment and functional outliers.}
\label{tab:supp_edu_outlier}
\begin{tabular}{@{}lrrr@{}}
\toprule
Specification & \(N\) & \(p\)-value & Partial functional \(R^2\) \\
\midrule
Education-adjusted & 57 & 0.3517 & 0.0166 \\
Outlier-trimmed, 3 trajectories excluded & 56 & 0.3401 & 0.0176 \\
\botrule
\end{tabular}
\footnotetext{The relative \(L^2\) shifts from M3 are 0.178 and 0.143, respectively. The education-adjusted model excludes the two identified participants whose educational attainment was reported in the residual ``Other'' category and could not be placed on the ordinal scale. The excluded outliers had financial-literacy scores \(1,2,2\); mean financial literacy was \(1.67\) among excluded participants and \(2.00\) among retained participants.}
\end{blocktable}

\begin{blocktable}
\captionof{table}{Near-flat sensitivity based on the largest raw fractional discount \(1-\min_t f_i(t)\).}
\label{tab:supp_nearflat}
\begin{tabular}{@{}rrrrrr@{}}
\toprule
\(\epsilon\) & Additional exclusions & \(N\) & \(p\)-value & Partial \(R^2\) & Relative \(L^2\) shift \\
\midrule
0.00 & 0 & 59 & 0.2578 & 0.0216 & 0.000 \\
0.01 & 0 & 59 & 0.2578 & 0.0216 & 0.000 \\
0.02 & 3 & 56 & 0.1345 & 0.0361 & 0.352 \\
0.05 & 5 & 54 & 0.2159 & 0.0275 & 0.215 \\
0.10 & 15 & 44 & 0.1672 & 0.0411 & 0.356 \\
\botrule
\end{tabular}
\footnotetext{Rows with no additional exclusion reproduce the primary analytical sample and therefore the primary test.}
\end{blocktable}

\begin{blocktable}
\captionof{table}{Sensitivity to the reconstruction-identification tolerance for \(k_i\).}
\label{tab:supp_k_tolerance}
\begin{tabular}{@{}rrrrr@{}}
\toprule
Identification tolerance & \(N\) & \(p\)-value & Partial \(R^2\) & Relative \(L^2\) shift vs. \(10^{-8}\) \\
\midrule
0 & 78 & 0.6896 & 0.0042 & 0.782 \\
\(10^{-10}\) & 59 & 0.2578 & 0.0216 & 0.000 \\
\(10^{-9}\) & 59 & 0.2578 & 0.0216 & 0.000 \\
\(10^{-8}\) & 59 & 0.2578 & 0.0216 & 0.000 \\
\(10^{-7}\) & 59 & 0.2578 & 0.0216 & 0.000 \\
\botrule
\end{tabular}
\footnotetext{The smallest reconstructed \(k_i\) in the identified sample is \(9.43\times10^{-7}\). Every positive tolerance in the grid therefore selects the same participants and reproduces the primary test; the constancy of the reported values reflects a gap in the observed distribution of \(k_i\) rather than independent replications.}
\end{blocktable}

\begin{blocktable}
\captionof{table}{Smoothing sensitivity for the primary numerical financial-literacy test.}
\label{tab:supp_smoothing}
\begin{tabular}{@{}llrrr@{}}
\toprule
Configuration & Description & \(N\) & \(p\)-value & Partial \(R^2\) \\
\midrule
S0 & Primary & 59 & 0.2578 & 0.0216 \\
S1 & Rougher response basis & 59 & 0.2630 & 0.0216 \\
S2 & Smoother response basis & 59 & 0.2669 & 0.0216 \\
S3 & Coarser monotone reconstruction & 59 & 0.4384 & 0.0132 \\
S4 & Finer monotone reconstruction & 59 & 0.2594 & 0.0214 \\
\botrule
\end{tabular}
\end{blocktable}

\begin{blocktable}
\captionof{table}{Financial-literacy test after additionally conditioning on \(\log(k_i)\).}
\label{tab:supp_logk_global}
\begin{tabular}{@{}lrrr@{}}
\toprule
Specification & \(N\) & \(p\)-value & Partial functional \(R^2\) \\
\midrule
FL \(\mid\) TSS + Age + \(\log(k_i)\) & 59 & 0.0465 & 0.0555 \\
\botrule
\end{tabular}
\footnotetext{The relative \(L^2\) shift of the financial-literacy coefficient relative to M3 is \(0.467\). The pointwise band for the M4 financial-literacy coefficient excludes zero over \(38\%\) of the domain.}
\end{blocktable}

\clearpage
\section{Complementary full-sample analysis using \(g_i(t)=-\log\widehat f_i(t)\)}\label{secA3}

This appendix reports the global tests, coefficient functions, and band summary for the complementary function-on-scalar model estimated on the full analytical sample (\(N=139\)), with the smoothed log-discount trajectory \(g_i(t)\) as response.

\begin{blocktable}
\captionof{table}{Conditional Freedman--Lane tests for the complementary full-sample \(g(t)\) response.}
\label{tab:supp_g_tests}
\begin{tabular}{@{}lrrr@{}}
\toprule
Test & \(N\) & \(p\)-value & Partial functional \(R^2\) \\
\midrule
Numerical FL score \(\mid\) TSS + Age & 139 & 0.1155 & 0.0166 \\
FL factor \(\mid\) TSS + Age & 139 & 0.1034 & 0.0297 \\
\botrule
\end{tabular}
\end{blocktable}

\begin{blockfigure}
\includegraphics[width=\textwidth]{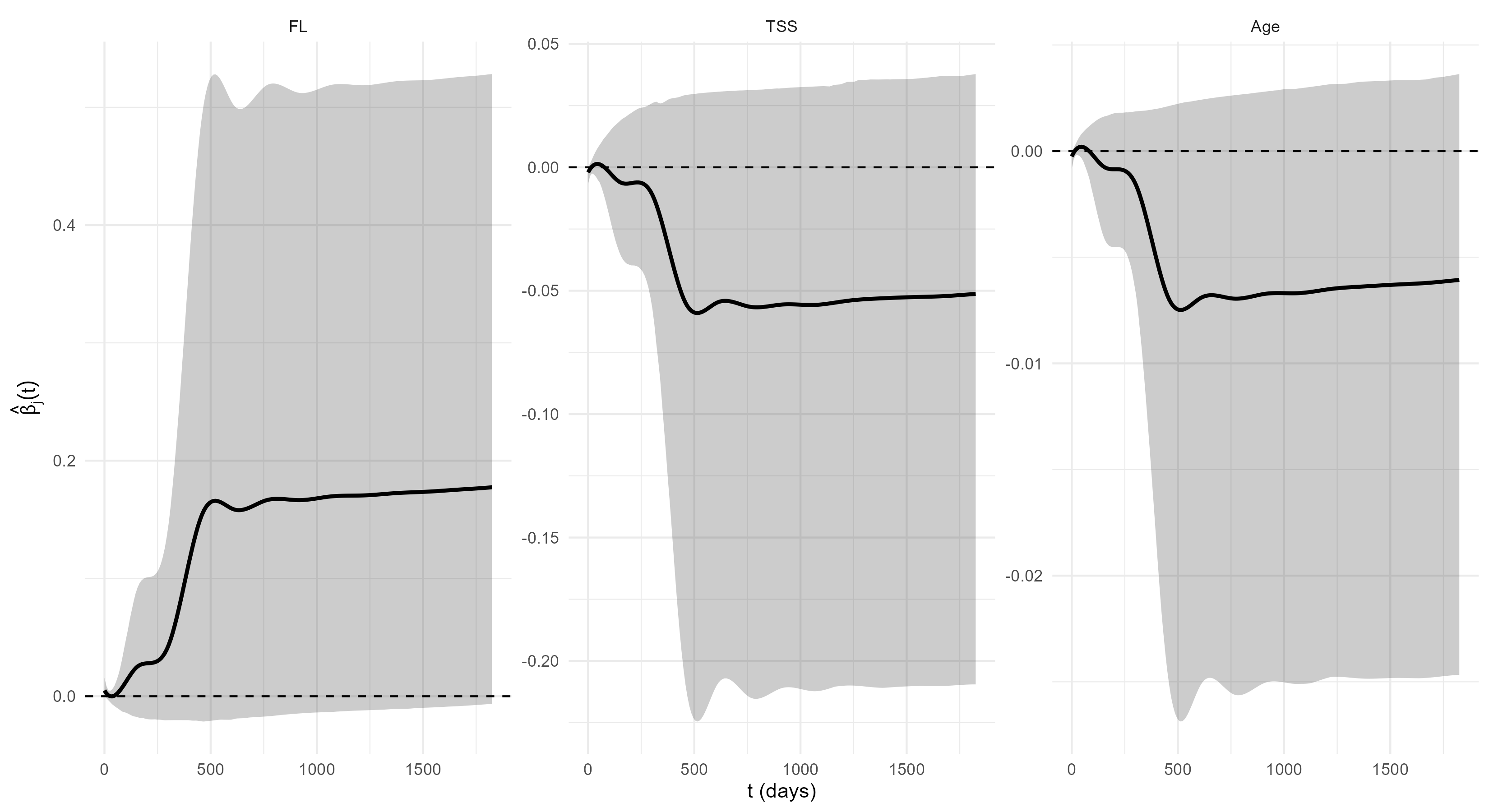}
\captionof{figure}{Coefficient functions for financial literacy, the Temporal Sense Scale, and age in the complementary full-sample \(g(t)\) model (\(N=139\)), with 95\% pointwise bootstrap bands.}
\label{fig:supp_g_beta}
\end{blockfigure}

\begin{blocktable}
\captionof{table}{Band summary for the numerical financial-literacy coefficient in the full-sample \(g(t)\) model.}
\label{tab:supp_g_bands}
\begin{tabular}{@{}lr@{}}
\toprule
Quantity & Value \\
\midrule
Simultaneous 95\% critical value & 2.344 \\
Share of domain with pointwise band excluding zero & 0.000 \\
Share of domain with simultaneous band excluding zero & 0.000 \\
\botrule
\end{tabular}
\end{blocktable}

\clearpage
\section{Measurement and smoothing diagnostics}\label{secA4}

This appendix reports the internal consistency of the Temporal Sense Scale and the effective degrees of freedom of the two monotone smoothing steps used in the reconstruction.

\begin{blocktable}
\captionof{table}{Internal consistency of the five-item Temporal Sense Scale in the full analytical sample.}
\label{tab:supp_tss_alpha}
\begin{tabular*}{0.5\textwidth}{@{\extracolsep{\fill}}lr@{}}
\toprule
Statistic & Value \\
\midrule
Number of items & 5 \\
Complete cases & 139 \\
Cronbach's \(\alpha\), raw & 0.879 \\
Cronbach's \(\alpha\), standardized & 0.879 \\
Mean inter-item correlation & 0.593 \\
\botrule
\end{tabular*}
\end{blocktable}

\begin{blocktable}
\captionof{table}{Effective degrees of freedom of the two monotone smoothing steps.}
\label{tab:supp_edf}
\begin{tabular}{@{}lrrr@{}}
\toprule
Step & \(n\) subjects & Mean edf & SD edf \\
\midrule
Discount-curve step (\texttt{mpd}) & 139 & 1.984 & 1.323 \\
Subjective-time step (\texttt{mpi}) & 59 & 4.845 & 1.566 \\
\botrule
\end{tabular}
\end{blocktable}

\FloatBarrier

\end{appendices}
\twocolumn

\FloatBarrier
\clearpage

\bibliography{references}

\end{document}